\documentclass[manuscript,screen, acmsmall]{acmart}
\AtBeginDocument{%
  \providecommand\BibTeX{{%
    \normalfont B\kern-0.5em{\scshape i\kern-0.25em b}\kern-0.8em\TeX}}}

\setcopyright{acmcopyright}
\copyrightyear{2022}
\acmYear{2022}
\acmDOI{https://doi.org/10.1145/3512904}

\acmBooktitle{Proceedings of the ACM on Human-Computer Interaction, Volume 6, CSCW1, Article 57, (November 2022), Taipei, Taiwan}
\acmPrice{15.00}
\acmISBN{978-1-4503-9999-9/18/06}

\usepackage{tabularx}
\usepackage{booktabs}
\usepackage{graphicx}
\usepackage{hyperref}

\usepackage{subcaption}
\usepackage{enumitem}

\usepackage{float} 
\usepackage{soul,color}
\usepackage[utf8]{inputenc}

\usepackage{longtable}
\usepackage[flushleft]{threeparttable}
\usepackage{chngcntr}
\usepackage{xcolor}
\definecolor{editCol}{rgb}{0.0, 0.0, 0.0}
\newcommand{\edit}[1]{{\textcolor{editCol}{#1}}}
\usepackage{pbox}
\usepackage{multirow}
\usepackage[colorinlistoftodos]{todonotes}
\usepackage{array}
\usepackage{makecell}
\usepackage{afterpage} 
\usepackage[title]{appendix}
\usepackage{subcaption,graphicx}
\usepackage{ragged2e}

\newcolumntype{L}[1]{>{\raggedright\let\newline\\\arraybackslash\hspace{0pt}}m{#1}}
\newcolumntype{C}[1]{>{\centering\let\newline\\\arraybackslash\hspace{0pt}}m{#1}}
\newcolumntype{R}[1]{>{\raggedleft\let\newline\\\arraybackslash\hspace{0pt}}m{#1}}

\begin{document}
\title[Joint Family Oversight for Mobile Online Safety and Privacy]{From Parental Control to Joint Family Oversight: Can Parents and Teens Manage Mobile Online Safety and Privacy as Equals?}

\author{Mamtaj Akter}
\email{mamtaj.akter@knights.ucf.edu}
\orcid{0000-0002-5692-9252}
\affiliation{%
  \institution{University of Central Florida, USA}
  \city{Orlando}
  \state{Florida}
  \postcode{32826}
}

\author{Amy J. Godfrey}
\email{amygodfrey@knights.ucf.edu}
\orcid{0000-0003-3825-5933}
\affiliation{%
  \institution{University of Central Florida, USA}
  \city{Orlando}
  \state{Florida}
  \postcode{32826}
}

\author{Jess Kropczynski}
\email{jess.kropczynski@uc.edu}
\orcid{0000-0002-7458-6003}
\affiliation{%
  \institution{University of Cincinnati, USA}
  \streetaddress{P.O. Box 1234}
  \city{Cincinnati}
  \state{OH}
  \postcode{45221}
}

\author{Heather R. Lipford}
\email{heather.lipford@uncc.edu}
\orcid{0000-0002-5261-0148}
\affiliation{%
  \institution{University of North Carolina, Charlotte, USA}
  \city{Charlotte}
  \state{North Carolina}
  \postcode{28223}
}
\author{Pamela J. Wisniewski}
\email{pamwis@ucf.edu}
\orcid{0000-0002-6223-1029}
\affiliation{%
  \institution{University of Central Florida, USA}
  \city{Orlando}
  \state{Florida}
  \postcode{32826}
}
\renewcommand{\shortauthors}{Mamtaj Akter et al.}

\begin{abstract}


Our research aims to \edit{highlight and} alleviate the \edit{complex} tensions around online safety, privacy, and smartphone usage in families so that parents and teens can work together to \edit{better} manage \edit{mobile privacy and security-related risks}. We \edit{developed a mobile application ("app")} for Community Oversight of Privacy and Security ("CO-oPS") \edit{and had parents and teens assess whether it} would be applicable for use with their families. CO-oPS is an Android \edit{app} that allows a group of users to co-monitor the apps installed on one another's devices and the privacy permissions granted to those apps. We conducted a study with 19 parent-teen (ages 13-17) pairs to understand how they currently managed mobile safety and app privacy within their family \edit{and then had them install, use,} and evaluate the CO-oPS app. We found that both parents and teens gave little consideration to online safety and privacy before installing new apps or granting privacy permissions. When using CO-oPS, participants liked how the \edit{app} increased transparency into one another's devices in a way that facilitated communication, but were less inclined to use features for in-app messaging or to hide apps from one another. \edit{Key} themes related to power imbalances between parents and teens surfaced that made co-management challenging. Parents were more open to collaborative oversight than teens, who felt that it was not their place to monitor their parents, even though \edit{both often believed parents} lacked the technological expertise to monitor \edit{themselves}. Our study sheds light on why collaborative practices for managing online safety and privacy within families may be beneficial but also quite difficult to implement in practice. We provide recommendations for overcoming these challenges based on the insights gained from our study.

\end{abstract}


\setcopyright{acmcopyright}
\acmJournal{PACMHCI}
\acmYear{2022} \acmVolume{6} \acmNumber{CSCW1} \acmArticle{57} \acmMonth{4} \acmPrice{15.00}\acmDOI{10.1145/3512904}


\maketitle
\section{Introduction}


Nearly every teen in the United States has access to a smartphone \cite{nw_teens_2018}, \edit{where they download and use numerous} applications ("apps") \cite{nw_teens_2013}. In some cases, these mobile apps facilitate risky online behaviors, like watching inappropriate content, texting with strangers, or sexting \cite{madigan_prevalence_2018}, which have heightened the concern of parents regarding the online safety of their children \cite{nw_parents_2016}. While concerned, parents are often unaware and underestimate the types and frequency of apps their teens use \cite{blackwell_managing_2016}. In an attempt to increase their awareness, some parents have turned to parental control apps to monitor their teens' smartphone usage and behaviors to keep them safe. However, teens often find these apps overly restrictive and invasive to their personal privacy \cite{ wisniewski_parental_2017, ghosh_safety_2018}. To resolve these tensions, researchers have called for more collaborative and teen-centric approaches where teens work alongside parents to make online safety decisions together \cite{cranor_parents_2014, hashish_involving_2014} and afford teens some level of personal privacy \cite{charalambous_privacy-preserving_2020, ghosh_circle_2020}. Apart from the online safety and privacy concerns related to parental control apps, third-party apps raise concerns for parents and teens alike, regarding sensitive information leakages \cite{reardon_50_2019}. Many mobile apps collect personal information (e.g., contact data, emails, photos, location, calendar events, and even browser history) \cite{nw_mobile_2015} from their users. As a result, a recent Pew Research study \cite{vogels_americans_2019} reported that the majority of U.S. adults have significant knowledge gaps about digital privacy and security. Ironically, younger generations are often more tech savvy than their older counterparts, so teens often provide technology support to their parents \cite{works_you_2019, correa_brokering_2015} when it comes to setting up and troubleshooting new digital devices \cite{kropczynski_examining_2021}. The paradoxical nature of parents wanting to protect their teens from online risks and privacy threats, while teens are more likely have more knowledge on how to achieve this, creates an interesting opportunity to design collaborative technologies where parents and teens could potentially support one another in accomplishing the shared goal of mobile online safety and privacy management within families.


Recent studies on adolescent online safety and parental mediation show that the influence between parents and teens is bi-directional and occurs in a more reactive 'just-in-time' fashion \cite{agha_just--time_2021}. Thus, technologies that provide real-time monitoring and promote bi-directional communication between parents and teens may be well-suited for supporting this dynamic. In a parallel but different research domain, networked privacy researchers (e.g., \cite{chouhan_co-designing_2019, aljallad_designing_2019, das_role_2015}) have been moving toward more collaborative and community-based approaches for co-managing mobile devices within trusted groups, including family members, friends, and co-workers. Yet, the parent-teen relationship is complex and nuanced in that adolescence is a unique developmental stage of seeking more independence from one's parents \cite{cranor_parents_2014}. Therefore, it is unclear whether community-based solutions for co-managing digital privacy and security would be generalizable to the parent-teen relationship. Therefore, \edit{we built and evaluated a mobile app for Community Oversight of Privacy and Security ("CO-oPS") and asked parents and teens to install, use, and evaluate the app} to \edit{assess whether a more collaborative solution for family privacy management would support parents and teens in online safety and privacy management, more so than one-sided parental control or tracking apps.} We designed our user study to answer the following high-level research questions:

 \begin{itemize}
   \item \textbf{RQ1:} \textit{How do parents and teens currently manage their mobile online safety and privacy?} 
   \item \textbf{RQ2:} \textit{As end users, how do they evaluate app features designed for co-managing mobile online safety and privacy?}
   \item \textbf{RQ3:} \textit{What are important considerations when designing an app for co-managing mobile online safety and privacy for families?}
 \end{itemize}

\edit{To answer these research questions, we conducted an in-depth user study with 19 parent-teen (ages 13-17) pairs. Participants were first asked about their current mobile privacy and security practices (RQ1). Then, we had them install and use the CO-oPS app to complete a set of guided tasks (RQ2). Finally, we asked them to reflect on whether the CO-oPS app would (or would not) be a good fit for the needs of their family (RQ3). Through our study, we disentangled how parents and teens conceptualized privacy and security differently in the context of mobile smartphone usage.} 


Overall, we found that most parents and teens made few considerations toward their own online safety or privacy when installing new apps or granting permissions to the apps they installed (RQ1). Meanwhile, parents often manually monitored the apps their teens installed but gave little thought to the permissions granted to those apps. Conversely, teens were less likely to care about the apps or permissions installed or granted on their parents' devices. When using CO-oPS (RQ2), parents and teens both liked features that gave them transparency into the apps and permissions of one another's devices, \edit{as they felt this increased transparency and would facilitate communication.} They also felt that while the app facilitated face-to-face conversations about safety and privacy, they did not need to use the in-app messaging features. Interestingly, parents and teens both disliked the privacy-oriented feature that allowed them to hide apps from one another. In terms of co-monitoring (RQ3), parents \edit{saw more value of joint oversight than teens.} Teens preferred the app for \edit{self-monitoring} because it raised their awareness, so they could more effectively manage the privacy and security of their devices themselves. Ironically, parents were more apt to take the advice given by their teens, while teens were more skeptical and would need to verify any advice given by parents prior to changing privacy settings or uninstalling apps at their parents' request. 


Our study makes important contributions to both the adolescent online safety and network privacy research communities by examining online safety, privacy, power, and trust when utilizing community-based oversight within families. \edit{Our research encourages the CSCW research community to think critically about the design of safety and privacy tools for families and demonstrates how online safety, privacy, and security are interrelated yet differing and complex concepts that surface unique tensions within the inherently hierarchical social relationships between parents and teens. Specifically, we make the following unique research contributions:}
 \begin{itemize}
   \item \edit{Through the novel design and use of the CO-oPS app, we elevated teens as equal partners to their parents in the co-managing of mobile privacy, security, and online safety}
   \item We gained empirical insights into the \edit{benefits but also the} power imbalance and tensions that make co-managing mobile online safety and privacy a challenge for these families; and
   \item We present considerations and design-based recommendations towards features that would better support parents and teens in providing joint oversight of one another.
 \end{itemize}

\section{Background}
\edit{We situate our research within the adolescent online safety and privacy literature and show how our work moves towards more collaborative approaches for co-managing security within families.}

\subsection{\edit{Adolescent Online Safety and Privacy Management within Families}}
A Pew research survey shows that more than half of parents in the U.S. are ``very concerned'' about their children's constant online presence ~\cite{nw_parents_2016}. As such, the rapid increase in smartphone usage among adolescents \cite{nw_teens_2018} has prompted researchers to study how mobile apps and internet access have affected teens' online safety \cite{madigan_prevalence_2018, nw_majority_2018}. \edit{What researchers have found is that parents are concerned because they lack direct knowledge of what their teens are doing on their mobile devices and, thus, underestimate their children's social media usage \cite{blackwell_managing_2016}}, which may lead to heightened risks (e.g., unwanted explicit content, harassment, or sexual solicitations) ~\cite{mitchell_trends_2014, calciati_automatically_nodate}. While concerned about online safety, parents and teens may also overlook privacy threats related to the information being shared via apps, rather than the social risks that these apps facilitate \cite{cranor_parents_2014}. \edit{While there are laws explicitly protecting children (e.g., the Child Online Privacy Protection Act) from unauthorized data breaches, researchers have found that many mobile apps violate these laws \cite{basu_copptcha_2020}. For instance, Feal \cite{feal_angel_2020} found that 72\% of parental control apps shared data with third parties without mentioning their presence in the apps' privacy policies.}  

\edit{To keep teens safe from online safety risks, some parents have turned to parental control apps; yet, the consensus among researchers (c.f., ~\cite{ali_betrayed_2020, ghosh_understanding_2016, wisniewski_parental_2017, ghosh_safety_2018, moser_parents_2017,pain_paranoid_2006}) is that these apps may not be effective in protecting teens online and can potentially harm the parent-teen relationship.} For instance, Pain observed that parents' constant surveillance and tracking through parental monitoring apps often caused paranoia and fear among teens \cite{pain_paranoid_2006}. To further investigate, Wisniewski et al. \cite{wisniewski_parental_2017} studied 75 commercially available parental control apps. They found that most traditional parental monitoring apps are overly restrictive and invasive to their teens' personal privacy. These parental control apps have supported more parental hierarchy and less of teens' autonomy and parent-teen communication. Due to these reasons, these apps have failed to foster positive parent-teen relationships. \edit{Instead, Wang et al. \cite{wang2021a} suggested that parents and children would prefer apps that provide transparency and feedback}, while Cranor et al. ~\cite{cranor_parents_2014} argued that teens should have some level of privacy in their online activities. \edit{Next, we synthesize the relevant literature that discusses the benefits of a more collaborative approach to family online safety.}

\subsection{Co-managing Online Safety and Privacy as a Family}
\edit{Several} studies have explained how \edit{adolescent} online safety apps could benefit from allowing parents and teens to work \edit{more collaboratively on protecting teens online.} Taking a joint approach focused on online safety, for instance, Hashish et al. \cite{hashish_involving_2014} \edit{proposed an app called ``We-Choose,'' which allowed  parents and children to work together in selecting which apps were appropriate for use. They found that this approach received higher levels of buy-in from children.} In a more recent study, Charalambous et al. \cite{charalambous_privacy-preserving_2020} proposed \edit{a ``Cybersafety Family Advice Suite'' (CFAS), where youth had a say in what online activities would be monitored by their parents to alert them of suspicious activity.} Further, Ghosh et al. ~\cite{ghosh_circle_2020} \edit{proposed ``Circle of Trust,'' a mobile app for text messaging that allowed teens the ability to negotiate trusted versus untrusted contacts with their parents, so that they were afforded a higher level of privacy when interacting with these individuals. A common theme among all of these approaches is a distinct paradigm shift to more collaborative approaches, where teens have some agency in how their parents monitor their mobile activities.}

\edit{While the adolescent online safety literature cited above has demonstrated the need for more collaborative approaches for helping parents and teens manage the online safety and privacy of teens, the networked privacy community has gone a step further to acknowledge that \textit{everyone} can benefit from joint oversight and community-based approaches for privacy and security management. For instance,} Vogels et al. \cite{vogels_americans_2019} found that users often face difficulty in managing their online safety and privacy; and therefore, often need others' advice. Rader et al. reported in their studies ~\cite{rader_stories_2012, rader_identifying_2015} that individuals often learn privacy strategies from their loved ones (e.g., families, friends, colleagues). Moreover, users are influenced by others' privacy behavior and adopt online safety tools to keep themselves safe online based on the advice of others ~\cite{mendel_susceptibility_2017, das_increasing_2014,das_role_2015}. \edit{In fact, teens often provide tech support within their families. Correa et al. ~\cite{correa_brokering_2015} found that teens are tech savvier than the adults within their families; and thus, often act as technology guides for them. Similarly, Kiesler et al. \cite{kiesler_troubles_2000} found teens often act as the `family guru,' the person in the family who others turn to for technical help. By bringing two disparate bodies of literature together, our work uncovers an apparent paradox; while the adolescent online safety literature relies heavily on parents to protect teens, the networked privacy literature acknowledges that teens are often technically more equipped to protect themselves online than their parents.}

\edit{Therefore, the novelty of our research contribution lies at the intersection of the adolescent online safety community's call for more collaborative approaches to teen online safety and the networked privacy community's call for more community-based approaches for helping \textit{all people} co-manage their digital privacy and security together. A novel aspect of the app we investigated is that it places parents and teens as equal partners in co-managing the apps installed and the privacy permissions granted to these apps on both parent and teen mobile devices. We are the first to move away from the unidirectional oversight of parents on teens' to the bidirectional oversight of parents and teens in managing mobile online safety and privacy. In the next section, we provide a detailed overview of the design of this app.}

\begin{figure}
\begin{subfigure}[t]{.295\textwidth}\centering
  \includegraphics[width=\columnwidth]{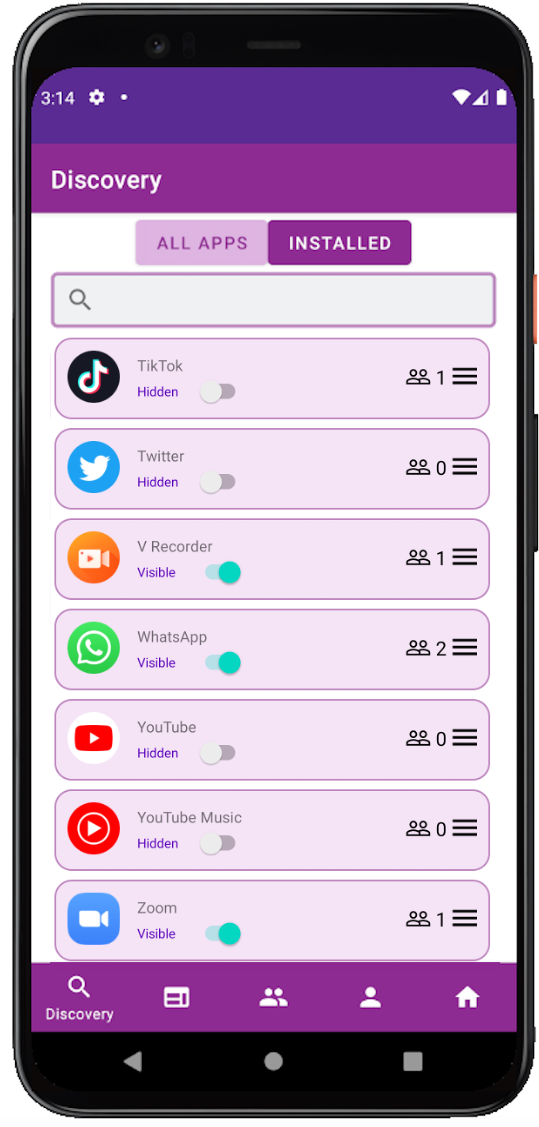}
  \caption{}
\end{subfigure}%
\begin{subfigure}[t]{.299\textwidth}\centering
  \includegraphics[width=\columnwidth]{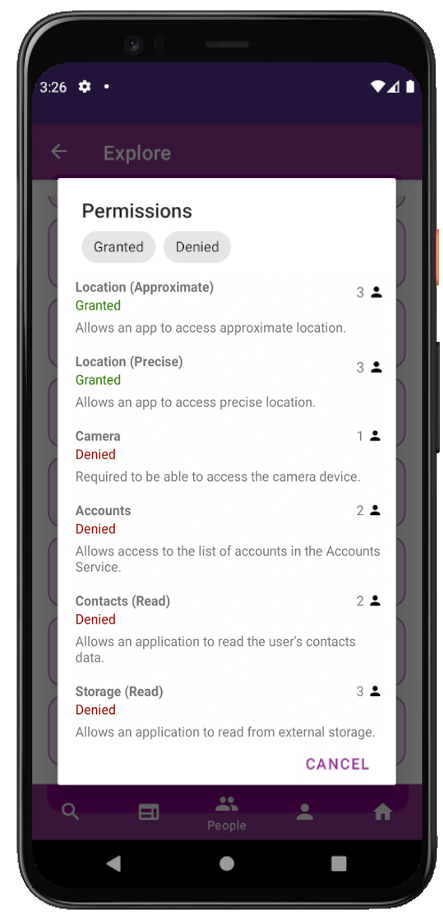}
  \caption{}
\end{subfigure}
\begin{subfigure}[t]{.295\textwidth}\centering
  \includegraphics[width=\columnwidth]{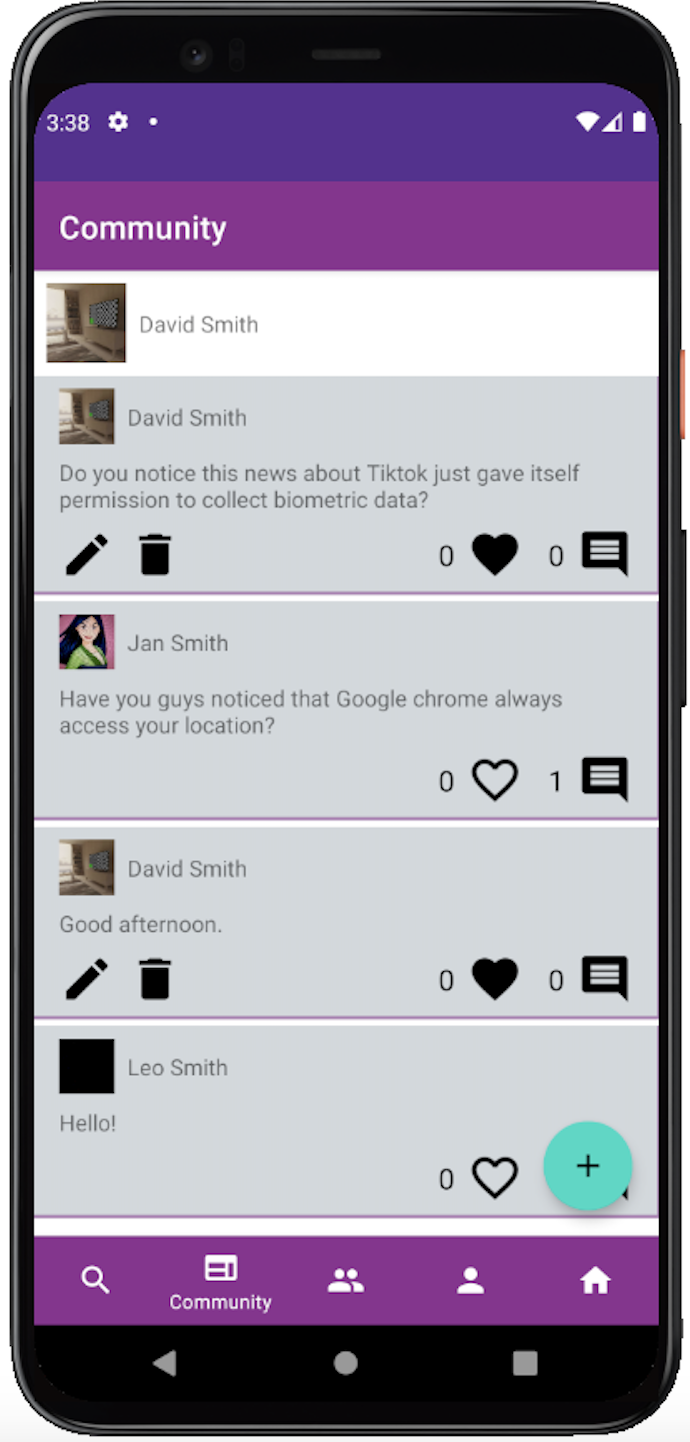}
  \caption{}
\end{subfigure}
\caption{\edit{CO-oPS Features: (a) App Hiding, (b) Collaborative Oversight, and (c) Communication}}~\label{fig:figure1}
\end{figure}

\section{Design of the CO-oPS App}
\edit{CO-oPS is based on Chouhan et al.'s ~\cite{chouhan_co-designing_2019} model of community oversight for privacy and security. This model proposes mechanisms aimed at enabling a community of trusted individuals to work together (through individual participation, transparency, awareness, and trust) to collectively manage their digital privacy and security. Chouhan et al. ~\cite{chouhan_co-designing_2019} interrogated their model through a participatory design study with groups of family members, friends, and co-workers. In a follow-up study, Kropcynski et al. \cite{kropczynski_towards_2021} further validated the applicability of the model of community oversight within older adult communities. Therefore, we adapted and implemented their initial app design \cite{aljallad_designing_2019}, which was instantiated from the model of community oversight to assess whether this approach would be suitable for the unique relationship between parents and teens.}

\edit{The CO-oPS app incorporated three key aspects: 1) App Hiding, 2) Collaborative App Management, and 3) Communication. Figure-\ref{fig:figure1}(a) shows an example of how users were able to turn on/off the visibility of their apps in the case they did not want them displayed to the other users. Next, users were able to collaboratively manage their privacy and security. For instance, there were given the ability to view what apps were installed on one another's mobile devices and what permissions were granted to each of these apps. Figure-\ref{fig:figure1}(b) presents an example screen shot of an app permissions. To promote communication, users were given the ability to post and comment on a Community Feed or Direct Message one another. Figure-\ref{fig:figure1}(c) shows an example screen shot of a Community Feed. Overall, these features supported a joint oversight and communication, allowing all community members to review one another's apps and share their feedback or guidance, while respecting each others' privacy.}

\section{Methods}
Below, we provide an overview of our study, the details regarding our data analysis approach, and then finally explains our recruitment strategy and the participants' profiles. 

\subsection{Study Overview}
\edit{Our user study consisted of three distinct phases: 1) Semi-structured interviews about how participants currently managed their mobile privacy and security, 2) A guided think aloud exploration of the CO-oPS app with probing questions, and 3) Semi-structured interview with parents and teens to reflect on whether the CO-oPS app would be a good fit (or not) for their family. We recruited 19 parent and teen pairs who completed all three phases of the research.} During the interview sessions, we had the parent-teen pairs install and use the CO-oPS app to perform guided tasks. Interview sessions took place on Zoom and were audio and video recorded. We then transcribed the recordings, and qualitatively coded for insights. For the first six participant dyads, parents and teens participated in the study together in the same Zoom session. However, we restructured the study to reduce groupthink bias \cite{noauthor_groupthink_2020}, so that subsequent parents and teens only came together for the final reflection (part 3). We facilitated this by having two researchers separately work with parents and teens in separate Zoom breakout rooms.

\edit{Table-\ref{tab:interview-questions} in Appendix A} presents sample interview questions organized by our overarching research questions. Each breakout session started with showing a storyboard of a third-party app requesting permission to access the user's phone storage \edit{to orient participants to the topic of the research. Then, we queried about privacy behaviors (Part-1) to better understand how they made decisions about installing and granting permissions to apps on their mobile devices. Next, we showed participants a video demo of the CO-oPS app to explain the core functionalities. The participants were then asked to perform a specific set of tasks using these different features, as shown in Table-\ref{tab:interview-questions} for RQ2. Throughout this part of the study, we encouraged the participants to think aloud \cite{guan_validity_2006} as they performed the tasks. We also asked contextualizing questions to better understand their experiences using the app. After the tasks were complete, parent-teen pairs exited their breakout rooms to take part in a group interview with the researchers. The study sessions were concluded by answering any final questions the participants may have had regarding our study. Then, we assisted them with uninstalling the app from their Android phones and emailed each parent-teen pair a \$20 Amazon gift card to thank them for their time. User study sessions took from one hour to two and a half hours to complete.}

\subsection{Data Analysis Approach}
First, we familiarized ourselves with our data by reading through each transcript. The first two authors under the supervision of the last author discussed the transcribed content to create our initial codes.  We identified several key dimensions that appeared to be the most influential during our analysis. Next, we conducted a grounded thematic analysis using Braun \& Clarke's \cite{braun2006using} six-phase framework to identify emergent themes and allow flexibility for new themes to emerge. The first author coded the parents' transcripts and the second author coded the teens' transcripts using the same set of initial codes, allowing for new codes to emerge. When a potential new code emerged, the researchers discussed whether they should be added to the codebook, and if they agreed, they went back to recode all prior transcripts to reflect that code. This was an iterative process, where the researchers constantly checked in with one another and formed a consensus. After all coding was complete, the first two authors worked with the last author to conceptually group the codes into cohesive themes that aligned with our over-arching research questions. Table-\ref{tab:codebook-RQ1}, \ref{tab:codebook-RQ2} and \ref{tab:codebook-RQ3} present the themes for each of our three research questions and their corresponding codes and illustrative quotations. Given the open-ended nature of the responses, participants, at times, gave conflicting answers. Therefore, we double-coded some responses, making the final count in these categories greater than the number of participants. For example, in our RQ2 codebook, \edit{parents and teens could identify multiple values in the collaborative privacy and online safety management. Therefore, the count of participants who thought a feature exhibited a particular value totaled to more than 100\% of our participants.} 

\subsection{Participants and Recruitment}
We recruited teens (age 13-17) and one of their parents or legal guardians (18 years or older). Parents were required to complete a \edit{pre-screening eligibility survey that verified whether they met the inclusion criteria of the study prior to providing their informed consent. Teens also provided informed assent to participate in the study.} The inclusion criteria for participation included: 1) Reside in the US and be fluent in English. 2) Both have Android smartphones and computers connected to the internet with Zoom access. 3) Be willing to install a beta app, CO-oPS, on their Android phones that collect \edit{the installed app names and their granted permissions}. We advertised through recruitment emails, phone calls, and by posting the flyers on social media. The recruitment process started in November 2020 and ended in April 2021. Overall, we recruited a sample of 38 participants (19 parent-teen pairs). \edit{Our study was Institutional Review Board approved.}  Table-\ref{tab:Demogrphic} shows the participant demographics. We had a diverse sample of 42\% Asian, 32\% Caucasian, 21\% Hispanic / Latino, and 5\% African American families. Among the teens, 53\% were females and 47\% were males; 58\% (N=11) of the parents were female and 43\% (N=8) were male. Teens were between 13 and 17 years old, with a mean age of 15.4 and a standard deviation of 1.4 years. The parents' age ranged from 40-55, with a mean age and standard deviation of 47.7 and 4.76, respectively. 

\begin{table}
  \centering
  \footnotesize
\caption{Participant Profiles}
  \label{tab:Demogrphic}
\begin{tabular}{ c c c c c c c} 
 \hline
 \textbf{Teen ID} & \textbf{Age} & \textbf{Sex}  & \textbf{Parent ID} & \textbf{Age} & \textbf{Sex} & \textbf{Ethnicity} \\ \hline
 \multicolumn{7}{c}{} \\[-10pt]  
 T1 & 17 & F  & P1 & 52 & F & White/Caucasian \\ 
 T2 & 15 & F  & P2 & 50 & F & White/Caucasian \\ 
 T3 & 14 & F  & P3 & 43 & M & Asian/Pacific Islander \\ 
 T4 & 17 & M  & P4 & 41 &  M  & Hispanic/Latino\\ 
 T5 & 16 & M  & P5 & 51 & F  & White/Caucasian\\ 
 
 T6 & 17 & M   & P6 & 54 & F & White/Caucasian \\ 
 T7 & 14 & F   &  P7 & 40 & M  & Asian/Pacific Islander \\ 
 T8 & 17 & M   & P8 & 46 & F  & Asian/Pacific Islander \\ 
 T9 & 16 & M   & P9 & 51 & F & White/Caucasian \\ 
 T10 & 17 & M  & P10 & 49 & M & Asian/Pacific Islander \\ 
 
 T11 & 13 & M   & P11 & 40 & F & Asian/Pacific Islander \\ 
 T12 & 16 & M   & P12 & 55 & M & Asian/Pacific Islander \\ 
 T13 & 13 & F   & P13 & 48 & F & Asian/Pacific Islander \\ 
 T14 & 16 & F   & P14 & 53 & M & Hispanic/Latino \\ 
 T15 & 15 & F   & P15 & 40 & F & Black/African American \\ 
 
 T16 & 14 & F   & P16 & 49 & M & Hispanic/Latino \\ 
 T17 & 14 & F   & P17 & 51 & F & White/Caucasian \\ 
 T18 & 17 & M   & P18 & 46 & F & Hispanic/Latino \\ 
 T19 & 16 & F   & P19 & 48 & M & Asian/Pacific Islander \\ \hline
\end{tabular}
\end{table}

\section{Results}
In our results, we use illustrative quotations to describe each of the themes that emerged from our qualitative analysis. Teen's quotations are identified by their IDs (e.g., T1,..T19), age, and gender information. Parents' quotations are identified by their IDs (e.g., P1,..P19) and teen's information. 


\subsection{Current Approaches for Managing Mobile Online Safety and Privacy (RQ1)}
This section presents parents and teens' current practices that they followed to manage their mobile online safety and digital privacy \edit{and sheds light on how they monitored one another's app safety.}


\subsubsection{Most parents and teens install apps with little consideration about privacy and online safety.}

Overall, we found that parents and teens had given little previous thought to mobile online safety and \edit{digital} privacy when they installed new apps. Most of the participants (63\%, N=12 Parents and 52\%, N=10 Teens) said that they took \textbf{no special steps} to verify an app's safety, but rather they installed apps on an as-needed or as-wanted basis, such as for school, work, entertainment, or other reasons. In many cases, the reason did not have to be well-thought-out, especially for teens, who were apt to install new apps just because they were bored:

\begin{quote}
\textit{
"It's more like, I kind of install it [Apps] maybe either school related or I'm just being bored and I will just install a game or something."} -T4, Male, 17 years
\end{quote} 

\noindent
However, about one-third of the participants (32\%, N=6 Parents and 37\%, N=7 Teens) said they would \textbf{do some research} before using an app. This was the case more so for teens than parents. For instance, they said they would look up the app online to see the number of downloads, or learn about the company that developed the app. They also reported that they read reviews on the Google Play Store and sometimes read articles online about the app before downloading any. 

\begin{quote}
\textit{"I usually try to look at their [Apps] reviews, to be honest, and do a little bit of background research on the app. So yeah. Just look it up on Google."} -T2, Female, 15 years
\end{quote} 

\noindent
About \edit{a quarter} of teens (26\%, N=5) reported that they would \textbf{seek their parents' permissions} to ask if they \edit{could} install an app or not. \edit{Only one} parent (P12) said that he would ask his kids about an app before he installed it.  

\begin{quote}
\textit{
"Usually I will talk to my son, or maybe daughter, if they know of this specific app, or if they have used this specific app before, then yeah, I talked to them first, before downloading them."} – P12, Father of T12 (Male, 16 years)
\end{quote} 

\noindent
Most participants took few precautions before installing new apps. But, teens showed more care to consider safety than their parents, as indicated by some seeking parents' permission before downloading an app. In the next section, we present their online safety behaviors centered around granting the app permissions. 

\subsubsection{Most parents and teens accepted permissions for their apps to function properly.}
In terms of how parents and teens decided on whether an app permission was safe to accept or not, we also found that most of the participants (79\%, N=15 parents and 89\%, N=17 teens) \edit{said they} would simply accept the app permission requests because \textbf{the apps otherwise would not work}. For example, a parent, P15, explained that some apps do not function properly \edit{otherwise.} In these cases, participants often \edit{said they} either decided not to install an app or grant the permissions, so that they could use the app for its intended purpose.

\begin{quotation}
\textit{"There are some apps that if you don't accept all permissions, it won't work properly, so either I may just choose not to use that one or just decide that I'm okay with the things its asking to me. So it really depends on if I need that app.”} –P15, Mother of T15 (Female, 15 years)
\end{quotation}


\noindent
There were a few parents (N=3, 16\%) and teens (N=2, 11\%) who said they would \textbf{always hit "accept"} because they were not aware or concerned about the potential consequences. 


\begin{quote}
\textit{
“To be honest, I never thought about it. I usually accept.”} – P8, Mother of T8 (Male, 17 years)
\end{quote} 

\noindent
Only one parent, P4, expressed how concerned he was about his mobile data privacy and therefore he would \textbf{deny all} the permission requests. Next, we discuss how the participants described engaging in monitoring their family members' online app safety. 


\subsubsection{Although teens provided general tech support to their parents, parents often manually check their teens’ app usage.}
Almost half of the parents (48\%, N=9) said they often \textbf{manually checked} their teen's apps usage. This meant that they physically took their teens' phones and would look through their installed apps or the general usage of their phones. Parents were often concerned about their teens' online safety because teens may not have reasonable boundaries on their actions; therefore, parents felt that checking their teens' phones is one of their parental responsibilities. Alongside checking teens' app usage, parents also mentioned that they go through the message contents that their teens exchange with their contacts, as described by one father:

\begin{quote}
\textit{
“Well, I do monitor those [Teen's phones] because that's what we [Parents] are for. I do see any kind of conversations she has with her new friend or anybody else, she has to talk to me, like I want to know what that conversation is. I want to know where it's taking place. Or I want to look at her text now and then. So everything else, you know, look at her videos that she's watching.  I kind of see what are the apps or tools she has on her phone.”} – P3, Father of T3 (Female, 14 years)
\end{quote} 

\noindent
About a quarter of parents (26\%, N=5 parents) reported that they \textbf{used parental control apps} on their teens' phones to support awareness of what their teens were using or doing. Most of these parents also mentioned that they use parental control apps because they could set the ground rules for their teens' online content access and also parents could be notified of their teens' online activities. The two most common parental control apps mentioned by parents are Google Family Link and Net Nanny, as described by one father below. 


\begin{quote}
\textit{“I have an app named Google family and I use that one. And there's another one that I use it for his game console. So that's pretty much it. I monitor through that. And then I put some kind of restrictions sometimes. There are certain stuffs that are not for him that are very adult contents that I cannot but block.”}-P4, Father of T4 (Male, 17 years)
\end{quote}

\noindent
The rest of the parents (26\%, N=5) \textbf{did not check} their teen's phone or apps because they trusted their teens to manage their own mobile online safety, especially if their teens were older. For example, one mother said of her son:

\begin{quote}
\textit{"I usually don't monitor what he does, like on the apps, because I kind of trust him to know his boundary and also because he is already 16, I know how I have raised him. So I usually don't feel the need to have to monitor him what he does on his phones and his apps."}-P5, Mother of T5 (Male, 16 years)
\end{quote} 


\noindent
In terms of monitoring parents' app safety, no teens mentioned they would check their parents' apps or phone usage. As a reason for not checking, teens mostly felt that parents had a separate level of personal privacy and freedom compared to teens, like T13:

\begin{quote}
\textit{
“I mean, for parents, it is so that they can use any apps as they need and they might not want us to see their phones, right? And they need their space, they would not want us to know what they are doing. ”} - T13, Female, 13 years
\end{quote} 
\noindent
Although teens did not monitor their parents' app usage, they were often the providers of tech support in their families. When parents and teens were discussing about who is the most tech savvy person in their family or who would they consult to regarding their app safety concerns, the pairs frequently mentioned general tech support that they received or provided to one another in their families. Most of the families (74\%, N=14) said the teens were the most tech savvy people in their families, and when they needed help managing their devices, teens \textbf{provided that tech support}. For example, one parent explained that her teen knew more about computers than she did, so she would go to him to seek help:
\begin{quote}
\textit{
“That would be my son. He is the one we all call when we're stuck with something. He then can look up. And he really finds out if it's not working properly.”} - P9, Mother of T9 (Male, 16)
\end{quote} 
\noindent
On the other hand, about a quarter of the families (26\%, N=5) reported that parents were the tech savvy ones and provided tech support to their teens. As such, other family members sought their help in troubleshooting app login issues, WiFi connections, or the phone and computer settings. 

\begin{quote}
\textit{
"I am, because in my family, I take care of the devices. So whenever they have any issues with their phones or computers, I try to fix that."} - P7, Father of T7 (Female, 14 years)
\end{quote} 


\noindent
\edit{In summary, our findings for RQ1 largely mirror prior work ~\cite{felt_android_2012, felt_ive_2012-1}, which found that general mobile phone users take little consideration regarding the apps installed and permissions granted on their smartphones. Yet, we also uncovered that almost all participants felt that they had little control over the decision to accept the permission requests given the need for the apps to function properly. Our work also confirmed that teens, not parents, were often viewed as the tech experts in the family.} 

\subsection{Feature Evaluation of CO-oPS App (RQ2)}

\edit{Based on participant interviews and the design of the CO-oPS app (Figure-\ref{fig:figure1}),} we divided the features into three broad categories: 1) Features for collaborative privacy and online safety management, 2) Features that facilitated communication, and, 3) Features that facilitated privacy. Participants, in general, positively evaluated the collaborative privacy and safety management features but disliked the feature that \edit{enhanced privacy through hiding apps}. Participants also did not see \edit{much} value in using the communication features for providing oversight or receiving feedback.

\subsubsection{Parents and teens valued features for collaborative privacy and online safety management.}
The features of collaborative privacy and safety management that allowed users to review one another's apps and app permissions were found to be the most popular features. Most of the families (63\%, N=12 Parents and 74\%, N=14 Teens) found these features beneficial in terms of \textbf{ensuring online safety}. \edit{Interestingly, teens often} talked about online safety and security synonymously. 

\begin{quote}
\textit{
"The benefit is you can warn family members from dangers. You can be able to do security check every now and then to make sure that they are not doing anything weird with the apps, or, you know, for keeping them safe."} -T5 (Male, 16 years)
\end{quote} 
\noindent
In contrast, parents cared less about security from a technical standpoint in favor of making sure their teens were protected from online dangers. Parents were mostly concerned about apps that facilitated unsafe social interactions or had a bad reputation of being an unsafe place for teens.

 \begin{quote}
     \textit{"It gives a piece of mind that I now know what apps they have and they are not doing anything unsafe, you know, I can track what apps my daughter and others are using."} -P7, Father of T7 (Female, 16 years)
 \end{quote} 

\noindent
Additionally, participants (68\%, N=13, of Parents and 42\%, N=8, of Teens) mentioned that these features \textbf{allowed joint oversight} where others could look over their apps and permissions and to point out any concerning apps or permissions they may have been using. Participants often mentioned that having another set of eyes would help them be less worried about not being aware of what concerning apps and granted permissions they had on their mobile phones. 

\begin{quote}
\textit{
“I like sort of the ability of a group to look at a security instead of individuals having to, you know, maintain security, having more than one set of eyes look at your apps is probably a good thing.”} -P3, Father of T3 (Female, 14 years)
\end{quote} 

\noindent
Some parents (37\%, N=7) also saw these features as a way to \textbf{promote awareness} among their families. They often explained how providing feedback on one another's apps and permissions could make everyone in their families more aware about online safety and privacy. 

\begin{quote}
\textit{
“Actually, I like it actually, this and the creating this awareness that they should not download and give permission, like as a blank check to all their apps. This awareness can be created among everyone.”} -P12, Father of T12 (Male, 16 years)
\end{quote} 

\noindent
Overall, all parents and teens found benefit in co-managing their app safety with one another. Next, we discuss how the participants found the communication features useful for their families. 

\subsubsection{The communication features were not appreciated among families. Parents preferred talking instead of messaging.}

The CO-oPS app had multiple forms of communication embedded within the platform for ease of communication. Like the features for collaborative privacy and online safety management, our initial assumption was that the communication features may also be advantageous to allow users to provide and receive feedback on apps and permissions. However, our participants expressed that \edit{their preferred mode of contact would be to talk to one another.}

When it came to using the CO-oPS communication features, over half of the parents (58\%, N=11) expressed interest in providing oversight in person or face-to-face, rather than using the communication features.  About a quarter of the teens (26\%, N=5) also said they \textbf{would rather talk} to their parents in person instead of using the instant messaging or the community feed of CO-oPS. 
  \begin{quote}
     \textit{
     "I think it would be better to discuss these types of issues in person. So I wouldn't really use this message feature."}-T7, Female, 14 years
 \end{quote} 
 
\noindent
Because the pairs \edit{lived in the same house}, some parents (37\%, N=7) and teens (21\%, N=4) mentioned that they would use these communication features \textbf{only when they were physically apart}.

 \begin{quote}
\textit{
 “If she [Parent] wasn't home, then yes, then I would use it [Communication feature].”} - T11, Male, 13 years
 \end{quote} 
 
\noindent
Some parents (21\%, N=4) and teens (32\%, N=6) said they would not use the communication features at all because they already \textbf{use other messaging tools} (e.g., texting from a messaging app) to communicate. Using a separate app would be redundant and unnecessary.

\begin{quote}
\textit{
 "I would probably just send a regular texts. It doesn't have to be through this [Communication feature]. Because that's where we primarily send our messages."}-P15, Mother of T15 (Female, 15 years)
 \end{quote} 
 
\noindent
While most of the families were reluctant to use the communication features, \edit{about a quarter} of parents (N=5, 26\%) and teens (N=4, 21\%) mentioned that these features could be useful to \textbf{initiate discussion} about app safety and data privacy. For example, parent P14 felt that this would help their entire family make better decisions, especially if she could use the feature to broadcast a potential threat to all her family members at once:

\begin{quote}
\textit{
 "Like if I read a news about an app or something, I would send community message probably to all, I'd say I'm concerned about an app. I will be probably cautioning everyone not to use that app because it's using something fishy accessing private information or something like that. That's how everyone could also take part in the discussion."} -P6, Mother of T6 (Male, 17 years)
 \end{quote} 
 
\noindent
Next, the parents and teens expressed their reactions to the feature that allowed them to hide any apps that they are not comfortable in sharing with their families.

\subsubsection{Both Parents and Teens did not approve of the features that decreased transparency.}
The feature that garnered the most discussion among parents and teens was the ability to hide or show apps to one another. Overall, parents and teens were both concerned about this feature because it promoted secrecy and negated some of the purpose behind the app. One important thing to be noted here is that when we asked teens about their opinion regarding this privacy feature, they mentioned that this feature could be harmful for kids, in general. 84\% of Parents (N=16) and 32\% of teens (N=6) mentioned that this feature may \textbf{promote dangerous behaviors}. For example, T10 commented on the risk of allowing teens to hide apps and said:
 \begin{quote}
\textit{
“If they're [Teens] using an app that's like, dangerous, and then they hide that, that could lead to some problems later on. Yeah, that could bring harm to them or to the entire family.”} -Teen 10, Male, 17 year
\end{quote} 

\noindent
We noticed an ambivalence among teens here. The teens who discussed the pros and cons of this feature \edit{in the breakout room} (separately from their parents) tended to be less negative about it than those who discussed it in front of their parents. Further, teens who raised concerns about the feature often talked using third-person "they" pronoun instead of collective "we" or first-person "I" statements. So, here, teens mostly did not talk from their own points of views, rather they talked in general from other teens' perspectives. For example, 

\begin{quote}
\textit{
"I would say, don't allow the hide feature. Because teens could just, it could be problematic if they have apps that they are hiding, it kind of compromises the whole purpose of the app [CO-oPS]. Because the majority of the time if people are hiding something, they're doing something wrong."} -T14, Female, 16 years
\end{quote} 

\noindent
Almost half of the participants (42\%, N=8 Parents and 53\%, N=10 Teens) pointed out that this feature may \textbf{affect the transparent relationship} in their families. They primarily believed in a bi-directional transparency-based relationship and therefore, also expect their teens/parents not to hide any apps from them. Many of the parents emphasized on the importance of teens being open with parents, especially when it comes to their online safety. For example, one of the parents said: 

\begin{quote}
\textit{
“In our family, we are very open. I wouldn't want to know that my teens might be keeping an app from me, but obviously, I wouldn't want to encourage any kind of situation that might cause them to keep things private from me, like, it's not a quality that I want to encourage in my family. I don't like things that would encourage my family to hide things from each other.”} -P6, Mother of T6 (Male, 17 years)
\end{quote} 

\noindent
Apart from the above concerns, N=5 (26\%) parents and N=8 (42\%) teens said they \textbf{did not want to hide apps} that they had installed on their devices. Most teens said their parents already knew what apps were installed on their phones. Similarly, parents said they would not use any app that they would not share with their teens. Therefore, they would not need to hide any apps from their teens. \edit{Since T1 and her mother were in the same Zoom discussion, they corroborated this point for us in the joint interview:}

 \begin{quote}
\textit{
"My mom reads the news a lot and she's not thrilled that I have TikTok, frankly, neither am I. But, you know, here we are. And they know my apps, they know I have it. So there's not much point in hiding it."} -T1, Female, 17 years \newline

\textit{"Right, I would agree with that. She probably is more savvy than I am about those kind of current apps. We trust her to make good decisions at this point. And um, no, I also can't imagine any way I would want to keep any of the apps hidden from my daughter."} -P1, Mother of T1 (Female, 17 years)
\end{quote} 

\noindent
About a quarter of teens (26\%, N=5) said these features might \textbf{cause distrust in their families} and create tension. They were also concerned their parents would still physically go through their phones, since this feature did not allow parents to see all the apps that teens had on their phones.

 \begin{quote}
\textit{
"I guess since you know, that you're [Teen] able to hide apps, from them [Parents] so they might want to check it out anyway. See, you can't hide it anyway.”} -T15, Female, 15 years
\end{quote} 

\noindent
Although most participants disapproved of the feature that allowed them to hide apps from one another, some (37\%, N=7 Parents and 47\%, N=9 Teens) mentioned a positive aspect of this feature. They identified that this feature \textbf{enabled users to have personal privacy} on their app usage and a sense of independence. For example, T16 said: 
\begin{quote}
\textit{
“Um, so like, an advantage would be like, you could download anything. And like, you could also hide it. So like, even if it [the app installed] was something weird, no one would know.  You could also have your privacy for apps that you know, you don't want people to know about.”} -T16, Female, 14 years
\end{quote}

\noindent
Overall, \edit{parents and teens shared mostly similar opinions about the feature of the CO-oPS app. In comparison, parents were somewhat more positive about the potential for joint oversight and less positive about the idea of hiding apps, than teens. Both parties preferred face-to-face or other forms of digital communication, rather than the in-app communication features.}

\subsection{Considerations for Designing an App for Co-managing Mobile Privacy and Online Safety for Families (RQ3)}
Next, we discuss the important considerations that one needs to take into account before designing an app that allows co-managing mobile privacy and online safety for families.  

\subsubsection{Parents were more concerned about teens’ app usage, while teens were more focused on the permissions of parents’ apps.}
In reviewing the apps installed and the permissions granted on one another's phones, parents were found to be more concerned about their teens' app usage but teens generally found more concerning privacy permissions on parents' phones. 74\% of the parents (N=14) found apps on their teen's phones that could cause concern but only 21\% of teens (N=4) identified concerning apps on their parent's phones. Among them, parents (63\%, N=12) were mostly concerned about \textbf{the unknown apps} that their teens had but parents did not know the purpose of the app or why it was needed. On the contrary, only N=2 (11\%) of teens reacted to their parents' apps that they had not heard of before. For instance, when looking through the teen's apps, the parent P2 reacted to a gaming app that she did not know about, whereas her teen T2 took her consent before installing that app but the parent forgot about it. Since T2 and her mother were in the same Zoom discussion, they corroborated this point for us in the joint interview:
\begin{quote}
    \textit{"I see things on your phone I've never seen before. I see Braino. what is Braino?"} - P2, Mother of T2 (Female, 15 years) \newline 

\textit{
"It’s a game, You know that I had it on my phone. I showed it to you before."} - T2, Female, 15 years
\end{quote} 

\noindent
The rest of the parents (37\%, N=7) identified some potentially \textbf{concerning apps that they already knew} that their teens had been using. Through their interaction with CO-oPS, they explicitly brought up the app names that could be a concern and discussed how these apps may harm their teens. They mostly were concerned about their teens having \textbf{social media apps} (21\%, N=4 Parents and 11\%, N=2 teens). Instagram, discord, Facebook and Snapchat were the most frequently mentioned social media apps. For instance, P7 said that his daughter used Instagram and explained why using this app may expose online dangers to his daughter:
 \begin{quote}
\textit{
"She [Teen] has Instagram because one of her school programs wanted her to have this account. But this might have some problems because if she keeps accepting unknown people's invitations."}-P7, Father of T7 (Female, 14 years)
\end{quote} 

\noindent
A few parents (16\%, N=3) were also surprised to see some \textbf{financial apps} on their teens phones. For instance parent P16 found Cash App, a mobile payment service app, on the teen's phone and thought this app was pre-installed on his daughter's new phone that they purchased recently. 
 \begin{quote}
\textit{
"I see Cash app. I need to check if it's coming with this phone itself or... No, I did not check that thing earlier. Those are kind of apps like you can pay someone."} -P16, Father of T16 (Female, 14 years)
\end{quote} 

\noindent
Only 11\% (N=2) of the teens identified some of the apps that their parent's had been using but teens knew that these apps had some privacy issues. For example, T6 expressed his concern about the Facebook app that his mother had on her phone and his mother P6 agreed with him.
 \begin{quote}
\textit{
"I can see she [Parent] has Facebook. That's [Facebook app] suspicious, I keep telling her to get off of Facebook."}-T6, Male, 17 years 
\newline
\textit{
"Yeah, I know I have the Facebook app. Um, he is like telling me to get off Facebook. He tells me not to go Twitter, he tells me all the dangers of these sites."}- P6, Mother of T6 (Male, 17 years)
\end{quote} 

\noindent
In terms of identifying concerning permissions on one another's phones, teens deemed more permissions granted as suspicious than their parents. More than half of the teens (58\%, N=11) \textbf{found some granted permissions} on their parents' phones that they thought might be a safety concern. On the flip side, only 32\%, N=6 of parents could find concerning permissions on their teens' phones. Most participants, especially parents, (26\%, N=5 parents and 32\%, N=6 teens) tended to react to the permissions granted for the approximate or precise location. 

\begin{quote}
\textit{
“She [Parent] has her location on, like a lot for pretty much every app, which I feel is kind of weird, because I don’t know why a Play Store needs her location. Her settings needs her location, which I guess could make sense, but not Play Store.”} -Teen 8, male, 15 years
\end{quote} 

\noindent
Some participants (5\%, N=1 Parents and 26\%, N=5 Teens) also identified other permissions (e.g., camera, SMS, storage, account) as sensitive data that their installed apps had access to. 

\begin{quote}
\textit{
"I think it's kind of weird that the Google Play stores are able to like read your text messages and stuff. It says granted, SMS read in it says allows an application to read SMS messages."} -T19, Father of T19 (Female, 16 years)
\end{quote} 

\noindent
In the next section, we discuss the parents and teens' feelings on providing oversight to one another.

\subsubsection{Parents would monitor their teens’ and other family member’s privacy and online safety whereas teens would mostly monitor their own apps.}
Parents were more interested in monitoring their family's privacy and online safety, yet teens said they would mostly check their own app safety instead. Most of the teens (74\%, N=14) would not want to monitor their parents' apps. They would instead be interested in \textbf{checking their own apps and permissions}. Teens often felt that their parents were more aware and capable of their own mobile online safety and privacy, and therefore they might not need others' oversight. 

\begin{quote}
\textit{
"I think I would use it [CO-oPS] to check my own permissions. Because I usually take advice from my parents, and I know that they wouldn't download something that isn't safe."} –T7, Female, 14 years
\end{quote} 

\noindent
Similar to the reluctance that teens showed in monitoring their parent's apps, some teens also showed no particular interest about other family members' app safety and thought that monitoring someone else's apps could be privacy invasive for them and therefore they would rather limit themselves in monitoring their own apps only.

\begin{quote}
\textit{
"Honestly, I don't usually really like to observe what other people might use what apps, because I kind of respect their privacy. I wouldn't use it for them, to be honest, but I would rather see using it for my own."}-T4, Male, 17 years
\end{quote} 

\noindent
Similar to these teens, there were some parents (37\%, N=7) who said they would rather monitor their own apps only, because they trusted their teens for having good judgement about their own online safety. Often times parents mentioned that they trusted their teens because teens were tech savvier than the parents and therefore, they thought that it was okay to rely on teens to have them make decisions on their own app safety and privacy. Hence, these parents mostly wanted to check their own apps and permissions instead. 

\begin{quote}
\textit{"My kids are smarter. I respect their judgment so far. They haven't let me down. So I'm probably not going to see their stuff. I think I'd be just doing mine. "}-P18, Mother of T18 (Male, 17 years)
\end{quote}

\noindent
Only a few teens (26\%, N=5) said they would review their parents' apps and permissions because they felt that their parents might not be tech-savvy enough to check their own app permissions. Therefore, they wanted to provide the oversight to ensure their parents' online safety.

\begin{quote}
\textit{
"I would do probably more towards my parents because like, they aren't as tech savvy, so they don't know as much. So I'd like to see what they're up to."}-T10, Male, 17 years
\end{quote} 

\noindent
In terms of using CO-oPS to monitor the rest of their family members' apps, most parents (68\%, N=13) responded positively that they \textbf{would like to monitor everyone's apps in their family}. One important phenomenon we noticed during the group discussion session is that parents often talked about checking the digital privacy of their families, while at the one-to-one session in the breakout room, they were more focused on the online safety issues (e.g., checking installed apps) on their teens' phones and did not focus much on the digital privacy issues (e.g., checking permissions). But as they continued their discussion regarding the online safety and privacy issues more with the researchers and with their teens, they seemed to be brainstorming on the consequences of data leakage and therefore, began to also discuss reviewing the privacy permissions. 

\begin{quote}
\textit{"I would mostly use it [CO-oPS] to check my family's apps because that's one way specifically to point at the permissions that they're allowing or they're not allowing to apps? Because, I don't know, most of the time, how they installed those apps, what they're allowing what, they're sharing what. Because I see a couple of things here by practicing with you guys. Because I can see what he should not be allowing access to it. Or allowing for information to share."}-P4, Father of T4 (Male, 17 years)
\end{quote}

\noindent
Overall, parents would like to monitor their teens and other family members. Teens in contrast, showed less interest in providing oversight to their parents or to their siblings. In the next section, we discuss the findings on how parents and teens reacted to the privacy and safety oversight they received from one another.

\subsubsection{Parents would listen to their teen’s oversight, but teens need to verify first.}
Parents overall said they \textbf{would listen to their teen’s input}, but teens were less receptive of their parents' feedback. Teens said they would need to verify the feedback first. N=17, (89\%) of parents would uninstall an app immediately or change permissions if their teens warned them. Only N=2 (11\%) of the parents \textbf{would need to verify first} by looking up online.

\begin{quote}
\textit{
"I think I would definitely go in and change the permission. If somebody sent it to me, then I would say, yeah, let me know about it, and then I'll change the permission.”} – P3, Father of T3 (Female, 14 years)
\end{quote}

\noindent
Conversely, one third of the teens (32\%, N=6) would take their parents’ feedback and immediately act accordingly. The rest of the teens (68\%, N=13) would need to verify first by crosschecking the apps and permissions on their phones. Teens also mostly said that they would look up the app online that their parents' would think is concerning.

\begin{quote}
\textit{“Because there could be some moments of miscommunication where one party thinks the app is damaging, and the other party knows it isn't in some cases."}-T1, Female, 17 years
\end{quote} 

\noindent
The next sections sheds light on the tension that we observed among the parents and teens because of the collaborative monitoring and the lenient privacy granting properties of CO-oPS. 

\subsubsection{Where teens liked to be treated as peers in co-monitoring with their parents’ app safety, but parents wanted more power.} 
Even though parents valued the ability for joint oversight, they were uncomfortable with the fact that the app put them at an equal level to their teens. They felt that the app should somehow account for the asymmetry in the parent-teen relationship. Mostly parents are the ones who brought up this matter and expressed that they wanted more power than their teens. For example, one-third of the parents (32\%, N=6) \textbf{wanted some control on teens' apps or permissions}. They didn't want to just be able to co-monitor, they wanted to be able to directly fix problems identified, such as removing a risky app or denying a granted privacy permission. For example, parent P9 expressed that she did not like CO-oPS because it did not allow her to change or edit her son's installed apps or permissions. 
\begin{quote}
\textit{
“I was puzzled when you [Interviewer] pointed out that I can't deny access. You know, when I looked at his [Teen's] apps, I couldn't actually turn on or off stuff [change Teen's app settings]. I could see what was there and what was happening, but I can't change it. So my gut reaction was like, Oh, I don't like that.”} -P9, Mother of T9 (Male, 16 years)
\end{quote} 


\noindent
Another power tension was revealed when we observed that parents (32\%, N=6) \textbf{wanted to be notified} whenever their teens installed any new apps or changed any of the app permissions. 
\begin{quote}
\textit{
“If my daughter installs the new app. And so if it can notify me so that this person has installed a new app and, these, these are the permissions of that app. ”} -P19, Father of T19 (Female, 16 year)
\end{quote} 

\noindent
Among all the parents who wanted more power in this collaborative way of managing their app safety, N=5 (26\%) of parents explicitly said that they did not like CO-oPS because it \textbf{treated them as peers} (CO-oPS provided teens and parents the same ability in monitoring one another's app safety and privacy). For instance, parent P15 said:

 \begin{quote}
\textit{
 “The drawback I would say is they [Teens] have the same ability that the parent has. So that's the part I don't like, as a parent.”} -P15, Mother of T15 (Female, 15 years)
 \end{quote} 

\noindent
Parents also did not like the bi-directional transparency of the app usage. They did not appreciate the fact that CO-oPS allowed their teens to have the same level of privacy in their app usage as they did. 68\% of the parents (N=13) explicitly said that they would \textbf{not want their teens to have the ability to hide apps}, but they want do want that privacy feature for themselves.

\begin{quote}
\textit{
"Okay, so for example, the benefits might be like, if any personal app is installed, then I may not want to share all that with all the family members. So it gives you some kind of privacy. But for a child, I think up until she's 18 there should be some control and there should not be an option for them to hide anything."}-P7, Father of T7 (Female, 14 years)
\end{quote} 

\noindent
To mitigate this tension for equal privacy power, some (32\%, N=6) parents \textbf{wanted the control} feature in CO-oPS to decide whether they would want to allow their teens to have privacy or not. 
\begin{quote}
\textit{
"I guess maybe allowing the parent like control of what the teen can do, like with visibility and stuff. So like allow them if basically giving them the permission to make their apps visible or not visible.
”} -P18, Mother of T18 (Male, 17 years)
\end{quote} 

\noindent 
On the contrary, many teens (64\%, N=12) liked the fact that they were treated as peers with their parents in co-monitoring because it provided them a sense of empowerment. Teens identified the benefits of this co-monitoring process that CO-oPS offers to engage both parents and teens in helping one another in managing their app safety and digital privacy and therefore they often mentioned that this app seemed different than the controlling apps (e.g., parental control apps) that they have known before. For instance, one of the teens T19 described how CO-oPS did not just one sidedly took care of their parents' needs, it rather addressed teens' perspectives as well by treating both parties as equals.

\begin{quote}
\textit{
"I was gonna say, like, I think the main reason I like this app [CO-oPS] than other ones, because it's collaborative, rather than just like entirely one sided, or it's just like the parents controlling everything. It's like we are equal.”} -T19, Female, 16 years
\end{quote} 

\noindent
Some teens (16\%, N=3) even wanted the ability to edit their parents' app settings, since parents may not have the technological expertise to change app permissions from their own settings.

 \begin{quote}
 \textit{
"Because you have to physically go into settings and change those, because they [Parents] may not know how to go into the place and they can't do it. If you're able to do it from here [CO-oPS], that would be a thing like I need in this situation."}-T5, Male, 16 years
 \end{quote} 
 
\noindent
In sum, parents seemed more habituated with the commonplace notion of maintaining a hierarchical relationship with their children, and so expected CO-oPS to give them that feelings of control and dominating power over their teens. \edit{Interestingly, all these same parents who did not like to be treated as equal to their teens, also reported that they manually check their teens' phones or use parental monitoring apps.} Hence, we \edit{noticed} the above tensions or concerns among them when they realized that they had the same level of co-monitoring power as their teens had. To resolve these issues, they wanted more control over their teens' apps and permissions through this CO-oPS app. Teens, on the other hand, had a fairly positive opinion towards this equal co-monitoring. In the next section, we discuss the implications of these findings.

\section{Discussion} 
In this section, we describe the implications of our findings in relation to prior work and provide design implications for implementing a collaborative mobile online safety app for families. 


\subsection{The Convergence of Online Safety, Privacy and Security \edit{within Families}}
\edit{Our study has important implications in terms of how parents and teens conceptualized online safety, privacy and security differently}. Generally, parents were most concerned about what apps their teens had installed on their phones and were less concerned about the privacy permissions of the apps. This was because parents saw apps as access points to their children, either socially (e.g., social media) or financially (e.g., financial apps). In contrast, teens were more likely to consider it unsafe to share sensitive data, such as location, with apps. In a sense, this demonstrated how parents showed a distrust of other people, while teens were more distrusting of the technology platforms themselves. The differences we saw in how parents and teens conceptualized safety, risk, and privacy may be due to parents being less naive about potentially malicious interactions with other people \cite{madigan_prevalence_2018, nw_teens_2015-1}, and teens being more tech-savvy about the malicious intent of the technologies themselves \cite{nw_teens_2013}. Yet, by combining aspects of both safety and privacy into one app, we saw a convergence of discussions between parents and teens, as they shared their concerns about online safety alongside thoughts about app permissions. Given these overlapping concerns, by combining safety and privacy co-monitoring in one app, we might be able to help parents and teens share and balance these concerns with each other. \edit{Therefore, a key take-away from our research is the potential for joint oversight to act as a mechanism to raise teens' awareness about online safety risks facilitated through apps, while raising parents' awareness of privacy and and security risks of the apps themselves.} Thus, we urge future researchers to consider opportunities where adolescent online safety research and networked privacy research can be merged \edit{to create novel interaction patterns that have the potential to improve both goals, simultaneously.}

\subsection{Implications for Collaborative Approaches that Support Adolescent Online Safety}
While past research has called for more collaborative approaches for managing the mobile online safety of adolescents \cite{hashish_involving_2014, charalambous_privacy-preserving_2020, ghosh_circle_2020}, our findings suggest that the idea of putting parents and teens on equal footing may be a push too provocative for the comfort level of many of our participants. 
Unlike Ghosh et al. \cite{ghosh_circle_2020}, who redesigned traditional parental control apps around a mechanism of trusted and untrusted contacts, we pushed the idea of collaborative safety management to having teens perform the same monitoring of their parents' devices as was being done for them. By making co-monitoring bi-directional (i.e., parents and teens monitoring each other), some parents felt uncomfortable relinquishing control, while some teens did not understand why they would even bother to monitor their parents at all. Thus, an important clarification might need to be made in the online safety literature that collaborative approaches might need to be centered around giving teens agency and control alongside their parents in managing their \textit{own} online safety, rather than that of their parents'. However, we also believe that this hesitance may be due to the entrenched status quo towards parental controls as a mechanism for teen safety \cite{wisniewski_parental_2017} and argue that it might be time to change this status quo. 


Khovanskaya et al. \cite{khovanskaya_reworking_2017} pointed out how user-centered designs typically focus on developing incremental technologies that reflect the current state of the world and reinforce the status quo, rather than breaking societal norms that go beyond (or even against) the status quo. Yet, by focusing on immediate concerns (e.g., teen privacy and agency in online safety) instead of long-term solutions (e.g., the self-efficacy to protect oneself), we might miss an important opportunity where online safety technologies could potentially act as tools that not only empower teens to  protect themselves online, but also elevates them to play a key role in the online safety and privacy of other family members (e.g., parents, younger siblings, extended family members). Prior research has already shown how teens often act as technology experts within the home who provide technology support to family members \cite{correa_brokering_2015, kiesler_troubles_2000, mendel_my_2019}; therefore, increasing their capacity and self-efficacy to also provide online safety and privacy support may not be as far fetched as one would think. Yet, this kind of transformative change is difficult to make; therefore, such efforts must be taken deliberately and thoughtfully. How might we design online safety technologies to accentuate the strengths \cite{badillo2020towards} (rather than the deficits) of adolescents when it comes to online safety and risks? How can we get buy-in from parents to foster these strengths in the context of trusting their teens to make good online safety and privacy choices for themselves and their family?

\subsection{Implications for Collaborative Privacy Management beyond Families}
Our results can provide important lessons for collaborative privacy management and oversight even beyond parents and teens. First, people may have differing concerns and focus areas when it comes to online safety, privacy, and security. As mentioned above, parents were more concerned with the safety of apps, while teens focused on the privacy of particular permissions. A larger community may have an even wider set of differing concerns. A collaborative app such as CO-oPS would allow for users to share and communicate those different concerns with each other, and could lead to greater awareness amongst a community of different kinds of privacy risks. However, different concerns may also lead users to not appreciate any feedback that is given, or think that their feedback to others would not be valued. For example, teens did not expect their parents needed help with their online safety, and were thus reluctant to provide such oversight. This reluctance of users to participate in oversight to others was a concern raised by Aljallad et al. \cite{aljallad_designing_2019} in their initial study of the CO-oPS idea. Another threat to participating in collaborative privacy management is \edit{having doubts} about the feedback given by others. For example, we observed teens \edit{doubting} the guidance of their parents, as they perceived their parents as less tech savvy. 

Our study also highlighted the different power structures and expectations for co-monitoring that exist within a family. As discussed above, parents and teens did not expect to be treated as equals. Parents already monitor teens' online safety, but not vice versa. There may be other kinds of relationships with existing oversight expectations, for example, between adults and older parents, that result in this imbalance of power and expectations. In addition, our participants saw little reason for teens to hide individual apps, although parents may want such a privacy mechanism. Teens may also still need certain privacy controls with other family members or trusted community members. These results imply that the features available to some users may need to be different than for other users. A larger group of users may consist of a variety of different relationships, with different social dynamics and oversight expectations, that would thus need to be supported in order to enable users to participate in ways that fit the dynamics of their relationships. Researchers need to examine the types of roles, relationships, and features that would be needed to support different groups of people in collaborative privacy management. However, supporting different relationship structures within one application may be very challenging and how to do so remains an important open question.

\subsection{Implications for Design}
Our study provides insight into the features and mechanisms that would be needed in a tool for parents and teens to participate in joint oversight of their online safety and privacy. Our findings suggest additional features that designers should consider in supporting parents' and teens' needs.

\textit{Change Notifications: }
Participants reported that they wanted to be informed about any changes that take place in their family members' apps and settings. For example, parents in particular, wanted to be notified whenever their teens installed a new app. Thus, mechanisms to keep users aware of app changes will be important. For them, more traditional notifications were desired.  Future research will be needed to examine desired controls around such notifications, such as frequency and intrusiveness of them, as well as privacy controls.

\textit{Promoting \edit{Communication, But No Need for Features:}}
Parents and teens did not see value in using the app as a means of communicating about privacy and online safety issues. \edit{This may be because the teens lived in the same house with their parents and due to the Covid pandemic at the time, they were not spending much time apart. Yet, they felt that the transparency and knowledge of being able to view one another's installed apps and app permissions was a beneficial way to facilitate communication outside of the app.} So while an app for parents and teens may not need direct communication mechanisms, features to encourage and promote offline discussion of privacy and online safety should still be incorporated. For example, an app could provide prompts with topics or points of discussion based on a user's activities, and provide guidance for parents or teens to bring up sensitive issues or raise questions with each other.

\textit{Learning about Online Safety and Privacy: }
One of our findings suggested that teens would not listen to the feedback that they received from their parents because they had doubts that parents would be able to provide \edit{accurate advice}. Chouhan et al. \cite{chouhan_co-designing_2019} found similar issues, as their participants expressed an interest in including external expert users (e.g., an IT professional), so that community members could get more dependable expert advice and guidance. Thus, designers should explore ways to provide additional knowledge or expertise to teens and parents. For example, the app could allow parents and teens to include an additional member in their family network who may be more tech-savvy. The app could also provide features for users to connect with other kinds of external resources, \edit{such as app reviews from a trusted source.}


\textit{Encouraging Teens' Participation: }
We found that teens were more inclined to check their own app safety and did not show as much of an interest in providing oversight to their parents and other family members. Thus, incentive mechanisms would need to focus in particular on motivating and engaging teens. Research should investigate which incentives would be most effective. For example, nudges could remind teens to provide oversight, and reward mechanisms \edit{such as badges or points} could incentivize different activities. A reward system could also enable teens to gain further privileges or privacy mechanisms within the app as they participate at greater levels. Another possibility is to include additional teens in the app, such as siblings or other close family members, to encourage them to provide oversight and feedback to each other, in addition to their parents. 

\subsection{Limitations and Future Research}
While our study provided us an insightful understanding of using a collaborative approach to online safety in families, we also recognize several limitations of our study. \edit{For instance, since we conducted a lab-based study, our findings are based on an initial exploration of the app features, rather than long-term usage. While participants were generally favorable of our app, we realize that demand characteristics could have biased their opinions. Therefore, we plan to conduct a field study where parents and teens (and possibly other family members) could interact with CO-oPS in a natural setting for an extended period. Another limitation, but also a strength, of our study is 68\% of our families were people of color with half being of Asian descent (primarily Indian). While the CSCW community urges us to be inclusive of diverse voices, Asian, Hispanic, and Black parents tend to be more restrictive in their parenting styles than Caucasian families ~\cite{jambunathan_parenting_2002, livingstone_risks_2011}, which may have led them to evaluate our app differently than if we studied predominantly white families. Thus, in future studies, we would want to assess individual differences, such as parenting styles \cite{steinberg_impact_1992}, to better understand how different kinds of families assess and use the app over time. The end goal would be to develop online safety and privacy apps that support a diverse range of families.}

\section{Conclusion}
\edit{Managing mobile privacy and online safety within families is hard. This is especially true when considering the unique developmental needs of teens and the complex social relationships and power imbalances between teens and their parents. Therefore, we examined whether placing parents and teens as equal partners in family online safety and privacy management would be a feasible and/or helpful endeavor. We conclude that increased bidirectional transparency between parents and teens may be able to promote knowledge and facilitate communication that would then promote joint learning, oversight, and safety. However, these positive outcomes would rely heavily on buy-in from both teens and parents on the idea that it is their job to watch over one another. Yet, this notion is a drastic paradigm shift from the traditional online safety measures of parental control. Only time will tell if we, as a society and as individuals, can make such a shift in how we think about online safety and privacy within families.}

\begin{acks}
We acknowledge the contributions of Nazmus Sakib Miazi, Nikko Osaka, Anoosh Hari and Ricardo Mangandi, who developed the CO-oPS application. We would also like to thank the parents and teens who participated in our study. This research was supported by the U.S. National Science Foundation under grants CNS-1844881, CNS-1814068, CNS-1814110, and CNS-1814439. Any opinion, findings, and conclusions or recommendations expressed in this material are those of the authors and do not necessarily reflect the views of the U.S. National Science Foundation.
\end{acks}

\bibliographystyle{ACM-Reference-Format}
\bibliography{bibliography}


\begin{thebibliography}{49}


\ifx \showCODEN    \undefined \def \showCODEN     #1{\unskip}     \fi
\ifx \showDOI      \undefined \def \showDOI       #1{#1}\fi
\ifx \showISBNx    \undefined \def \showISBNx     #1{\unskip}     \fi
\ifx \showISBNxiii \undefined \def \showISBNxiii  #1{\unskip}     \fi
\ifx \showISSN     \undefined \def \showISSN      #1{\unskip}     \fi
\ifx \showLCCN     \undefined \def \showLCCN      #1{\unskip}     \fi
\ifx \shownote     \undefined \def \shownote      #1{#1}          \fi
\ifx \showarticletitle \undefined \def \showarticletitle #1{#1}   \fi
\ifx \showURL      \undefined \def \showURL       {\relax}        \fi
\providecommand\bibfield[2]{#2}
\providecommand\bibinfo[2]{#2}
\providecommand\natexlab[1]{#1}
\providecommand\showeprint[2][]{arXiv:#2}

\bibitem[\protect\citeauthoryear{??}{noa}{2020}]%
        {noauthor_groupthink_2020}
 \bibinfo{year}{2020}\natexlab{}.
\newblock \bibinfo{title}{Groupthink}.
\newblock
\newblock
\urldef\tempurl%
\url{https://en.wikipedia.org/w/index.php?title=Groupthink&oldid=991824085}
\showURL{%
\tempurl}
\newblock
\shownote{Page Version ID: 991824085.}


\bibitem[\protect\citeauthoryear{Agha, Ghaiumy~Anaraky, Badillo-Urquiola,
  McHugh, and Wisniewski}{Agha et~al\mbox{.}}{2021}]%
        {agha_just--time_2021}
\bibfield{author}{\bibinfo{person}{Zainab Agha}, \bibinfo{person}{Reza
  Ghaiumy~Anaraky}, \bibinfo{person}{Karla Badillo-Urquiola},
  \bibinfo{person}{Bridget McHugh}, {and} \bibinfo{person}{Pamela Wisniewski}.}
  \bibinfo{year}{2021}\natexlab{}.
\newblock \showarticletitle{‘{Just}-in-{Time}’ {Parenting}: {A}
  {Two}-{Month} {Examination} of the {Bi}-directional {Influences} {Between}
  {Parental} {Mediation} and {Adolescent} {Online} {Risk} {Exposure}}. In
  \bibinfo{booktitle}{\emph{{HCI} for {Cybersecurity}, {Privacy} and {Trust}}}
  \emph{(\bibinfo{series}{Lecture {Notes} in {Computer} {Science}})},
  \bibfield{editor}{\bibinfo{person}{Abbas Moallem}} (Ed.).
  \bibinfo{publisher}{Springer International Publishing},
  \bibinfo{address}{Cham}, \bibinfo{pages}{261--280}.
\newblock
\showISBNx{978-3-030-77392-2}
\urldef\tempurl%
\url{https://doi.org/10.1007/978-3-030-77392-2_17}
\showDOI{\tempurl}


\bibitem[\protect\citeauthoryear{Ali, Elgharabawy, Duchaussoy, Mannan, and
  Youssef}{Ali et~al\mbox{.}}{2020}]%
        {ali_betrayed_2020}
\bibfield{author}{\bibinfo{person}{Suzan Ali}, \bibinfo{person}{Mounir
  Elgharabawy}, \bibinfo{person}{Quentin Duchaussoy}, \bibinfo{person}{Mohammad
  Mannan}, {and} \bibinfo{person}{Amr Youssef}.}
  \bibinfo{year}{2020}\natexlab{}.
\newblock \showarticletitle{Betrayed by the {Guardian}: {Security} and
  {Privacy} {Risks} of {Parental} {Control} {Solutions}}. In
  \bibinfo{booktitle}{\emph{Annual {Computer} {Security} {Applications}
  {Conference}}} \emph{(\bibinfo{series}{{ACSAC} '20})}.
  \bibinfo{publisher}{Association for Computing Machinery},
  \bibinfo{address}{New York, NY, USA}, \bibinfo{pages}{69--83}.
\newblock
\showISBNx{978-1-4503-8858-0}
\urldef\tempurl%
\url{https://doi.org/10.1145/3427228.3427287}
\showDOI{\tempurl}


\bibitem[\protect\citeauthoryear{Aljallad, Guo, Chouhan, LaPerriere,
  Kropczynski, Wisnewski, and Lipford}{Aljallad et~al\mbox{.}}{2019}]%
        {aljallad_designing_2019}
\bibfield{author}{\bibinfo{person}{Zaina Aljallad}, \bibinfo{person}{Wentao
  Guo}, \bibinfo{person}{Chhaya Chouhan}, \bibinfo{person}{Christy LaPerriere},
  \bibinfo{person}{Jess Kropczynski}, \bibinfo{person}{Pamela Wisnewski}, {and}
  \bibinfo{person}{Heather Lipford}.} \bibinfo{year}{2019}\natexlab{}.
\newblock \bibinfo{title}{Designing a {Mobile} {Application} to {Support}
  {Social} {Processes} for {Privacy} ({Journal} {Article}) {\textbar} {DOE}
  {PAGES}}.
\newblock
\newblock
\urldef\tempurl%
\url{https://par.nsf.gov/biblio/10097722}
\showURL{%
\tempurl}


\bibitem[\protect\citeauthoryear{Anderson}{Anderson}{2015}]%
        {nw_mobile_2015}
\bibfield{author}{\bibinfo{person}{Monica Anderson}.}
  \bibinfo{year}{2015}\natexlab{}.
\newblock \bibinfo{title}{Mobile apps, privacy and permissions: 5 key
  takeaways}.
\newblock
\newblock
\urldef\tempurl%
\url{https://www.pewresearch.org/fact-tank/2015/11/10/key-takeaways-mobile-apps/}
\showURL{%
\tempurl}


\bibitem[\protect\citeauthoryear{Anderson}{Anderson}{2016}]%
        {nw_parents_2016}
\bibfield{author}{\bibinfo{person}{Monica Anderson}.}
  \bibinfo{year}{2016}\natexlab{}.
\newblock \bibinfo{title}{Parents, {Teens} and {Digital} {Monitoring}}.
\newblock
\newblock
\urldef\tempurl%
\url{https://www.pewresearch.org/internet/2016/01/07/parents-teens-and-digital-monitoring/}
\showURL{%
\tempurl}


\bibitem[\protect\citeauthoryear{Anderson and Jiang}{Anderson and
  Jiang}{2018}]%
        {nw_teens_2018}
\bibfield{author}{\bibinfo{person}{Monica Anderson} {and}
  \bibinfo{person}{Jingjing Jiang}.} \bibinfo{year}{2018}\natexlab{}.
\newblock \bibinfo{title}{Teens, {Social} {Media} \& {Technology} 2018}.
\newblock
\newblock
\urldef\tempurl%
\url{https://www.pewresearch.org/internet/2018/05/31/teens-social-media-technology-2018/}
\showURL{%
\tempurl}


\bibitem[\protect\citeauthoryear{Badillo-Urquiola, Agha, Akter, and
  Wisniewski}{Badillo-Urquiola et~al\mbox{.}}{2020}]%
        {badillo2020towards}
\bibfield{author}{\bibinfo{person}{Karla Badillo-Urquiola},
  \bibinfo{person}{Zainab Agha}, \bibinfo{person}{Mamtaj Akter}, {and}
  \bibinfo{person}{Pamela Wisniewski}.} \bibinfo{year}{2020}\natexlab{}.
\newblock \showarticletitle{Towards Assets-based Approaches for Adolescent
  Online Safety}. In \bibinfo{booktitle}{\emph{Badillo-Urquiola, Agha, Z.,
  Akter, K., Wisniewski, P.,(2020)“Towards Assets-Based Approaches for
  Adolescent Online Safety” Extended Abstract presented at the ACM Conference
  on Computer-Supported Cooperative Work Workshop on Operationalizing an
  Assets-Based Design of Technology,(CSCW 2020)}}.
\newblock


\bibitem[\protect\citeauthoryear{Basu, Hussain, Gupta, and Karri}{Basu
  et~al\mbox{.}}{2020}]%
        {basu_copptcha_2020}
\bibfield{author}{\bibinfo{person}{Kanad Basu}, \bibinfo{person}{Suha~Sabi
  Hussain}, \bibinfo{person}{Ujjwal Gupta}, {and} \bibinfo{person}{Ramesh
  Karri}.} \bibinfo{year}{2020}\natexlab{}.
\newblock \showarticletitle{{COPPTCHA}: {COPPA} {Tracking} by {Checking}
  {Hardware}-{Level} {Activity}}.
\newblock \bibinfo{journal}{\emph{IEEE Transactions on Information Forensics
  and Security}}  \bibinfo{volume}{15} (\bibinfo{year}{2020}),
  \bibinfo{pages}{3213--3226}.
\newblock
\showISSN{1556-6013, 1556-6021}
\urldef\tempurl%
\url{https://doi.org/10.1109/TIFS.2020.2983287}
\showDOI{\tempurl}


\bibitem[\protect\citeauthoryear{Blackwell, Gardiner, and
  Schoenebeck}{Blackwell et~al\mbox{.}}{2016}]%
        {blackwell_managing_2016}
\bibfield{author}{\bibinfo{person}{Lindsay Blackwell}, \bibinfo{person}{Emma
  Gardiner}, {and} \bibinfo{person}{Sarita Schoenebeck}.}
  \bibinfo{year}{2016}\natexlab{}.
\newblock \showarticletitle{Managing {Expectations}: {Technology} {Tensions}
  among {Parents} and {Teens}}. In \bibinfo{booktitle}{\emph{Proceedings of the
  19th {ACM} {Conference} on {Computer}-{Supported} {Cooperative} {Work} \&
  {Social} {Computing}}} \emph{(\bibinfo{series}{{CSCW} '16})}.
  \bibinfo{publisher}{Association for Computing Machinery},
  \bibinfo{address}{New York, NY, USA}, \bibinfo{pages}{1390--1401}.
\newblock
\showISBNx{978-1-4503-3592-8}
\urldef\tempurl%
\url{https://doi.org/10.1145/2818048.2819928}
\showDOI{\tempurl}


\bibitem[\protect\citeauthoryear{Braun and Clarke}{Braun and Clarke}{2006}]%
        {braun2006using}
\bibfield{author}{\bibinfo{person}{Virginia Braun} {and}
  \bibinfo{person}{Victoria Clarke}.} \bibinfo{year}{2006}\natexlab{}.
\newblock \showarticletitle{Using thematic analysis in psychology}.
\newblock \bibinfo{journal}{\emph{Qualitative research in psychology}}
  \bibinfo{volume}{3}, \bibinfo{number}{2} (\bibinfo{year}{2006}),
  \bibinfo{pages}{77--101}.
\newblock


\bibitem[\protect\citeauthoryear{Calciati, Gorla, Kuznetsov, and
  Zeller}{Calciati et~al\mbox{.}}{[n.d.]}]%
        {calciati_automatically_nodate}
\bibfield{author}{\bibinfo{person}{Paolo Calciati}, \bibinfo{person}{Alessandra
  Gorla}, \bibinfo{person}{Konstantin Kuznetsov}, {and}
  \bibinfo{person}{Andreas Zeller}.} \bibinfo{year}{[n.d.]}\natexlab{}.
\newblock \showarticletitle{Automatically {Granted} {Permissions} in {Android}
  apps}.
\newblock
\urldef\tempurl%
\url{https://doi.org/10.1145/3379597.3387469}
\showDOI{\tempurl}


\bibitem[\protect\citeauthoryear{Charalambous, Papagiannis, Papasavva,
  Leonidou, Constaninou, Terzidou, Christophides, Nicolaou, Theofanis,
  Kalatzantonakis, and Sirivianos}{Charalambous et~al\mbox{.}}{2020}]%
        {charalambous_privacy-preserving_2020}
\bibfield{author}{\bibinfo{person}{Markos Charalambous},
  \bibinfo{person}{Petros Papagiannis}, \bibinfo{person}{Antonis Papasavva},
  \bibinfo{person}{Pantelitsa Leonidou}, \bibinfo{person}{Rafael Constaninou},
  \bibinfo{person}{Lia Terzidou}, \bibinfo{person}{Theodoros Christophides},
  \bibinfo{person}{Pantelis Nicolaou}, \bibinfo{person}{Orfeas Theofanis},
  \bibinfo{person}{George Kalatzantonakis}, {and} \bibinfo{person}{Michael
  Sirivianos}.} \bibinfo{year}{2020}\natexlab{}.
\newblock \showarticletitle{A {Privacy}-{Preserving} {Architecture} for the
  {Protection} of {Adolescents} in {Online} {Social} {Networks}}.
\newblock \bibinfo{journal}{\emph{arXiv:2007.12038 [cs]}} (\bibinfo{date}{July}
  \bibinfo{year}{2020}).
\newblock
\urldef\tempurl%
\url{http://arxiv.org/abs/2007.12038}
\showURL{%
\tempurl}
\newblock
\shownote{arXiv: 2007.12038.}


\bibitem[\protect\citeauthoryear{Chouhan, LaPerriere, Aljallad, Kropczynski,
  Lipford, and Wisniewski}{Chouhan et~al\mbox{.}}{2019}]%
        {chouhan_co-designing_2019}
\bibfield{author}{\bibinfo{person}{Chhaya Chouhan}, \bibinfo{person}{Christy~M.
  LaPerriere}, \bibinfo{person}{Zaina Aljallad}, \bibinfo{person}{Jess
  Kropczynski}, \bibinfo{person}{Heather Lipford}, {and}
  \bibinfo{person}{Pamela~J. Wisniewski}.} \bibinfo{year}{2019}\natexlab{}.
\newblock \showarticletitle{Co-designing for {Community} {Oversight}: {Helping}
  {People} {Make} {Privacy} and {Security} {Decisions} {Together}}.
\newblock \bibinfo{journal}{\emph{Proceedings of the ACM on Human-Computer
  Interaction}} \bibinfo{volume}{3}, \bibinfo{number}{CSCW}
  (\bibinfo{date}{Nov.} \bibinfo{year}{2019}), \bibinfo{pages}{1--31}.
\newblock
\showISSN{2573-0142, 2573-0142}
\urldef\tempurl%
\url{https://doi.org/10.1145/3359248}
\showDOI{\tempurl}


\bibitem[\protect\citeauthoryear{Correa, Straubhaar, Chen, and Spence}{Correa
  et~al\mbox{.}}{2015}]%
        {correa_brokering_2015}
\bibfield{author}{\bibinfo{person}{Teresa Correa}, \bibinfo{person}{Joseph
  Straubhaar}, \bibinfo{person}{Wenhong Chen}, {and} \bibinfo{person}{Jeremiah
  Spence}.} \bibinfo{year}{2015}\natexlab{}.
\newblock \bibinfo{title}{Brokering new technologies: {The} role of children in
  their parents’ usage of the internet - {Teresa} {Correa}, {Joseph} {D}
  {Straubhaar}, {Wenhong} {Chen}, {Jeremiah} {Spence}, 2015}.
\newblock
\newblock
\urldef\tempurl%
\url{https://journals.sagepub.com/doi/10.1177/1461444813506975}
\showURL{%
\tempurl}


\bibitem[\protect\citeauthoryear{Cranor, Durity, Marsh, and Ur}{Cranor
  et~al\mbox{.}}{2014}]%
        {cranor_parents_2014}
\bibfield{author}{\bibinfo{person}{Lorrie~Faith Cranor},
  \bibinfo{person}{Adam~L. Durity}, \bibinfo{person}{Abigail Marsh}, {and}
  \bibinfo{person}{Blase Ur}.} \bibinfo{year}{2014}\natexlab{}.
\newblock \showarticletitle{Parents’ and {Teens}’ {Perspectives} on
  {Privacy} {In} a {Technology}-{Filled} {World}}. \bibinfo{pages}{19--35}.
\newblock
\showISBNx{978-1-931971-13-3}
\urldef\tempurl%
\url{https://www.usenix.org/conference/soups2014/proceedings/presentation/cranor}
\showURL{%
\tempurl}


\bibitem[\protect\citeauthoryear{Das, Kramer, Dabbish, and Hong}{Das
  et~al\mbox{.}}{2014}]%
        {das_increasing_2014}
\bibfield{author}{\bibinfo{person}{Sauvik Das}, \bibinfo{person}{Adam~D.I.
  Kramer}, \bibinfo{person}{Laura~A. Dabbish}, {and} \bibinfo{person}{Jason~I.
  Hong}.} \bibinfo{year}{2014}\natexlab{}.
\newblock \showarticletitle{Increasing {Security} {Sensitivity} {With} {Social}
  {Proof}: {A} {Large}-{Scale} {Experimental} {Confirmation}}. In
  \bibinfo{booktitle}{\emph{Proceedings of the 2014 {ACM} {SIGSAC} {Conference}
  on {Computer} and {Communications} {Security}}} \emph{(\bibinfo{series}{{CCS}
  '14})}. \bibinfo{publisher}{Association for Computing Machinery},
  \bibinfo{address}{Scottsdale, Arizona, USA}, \bibinfo{pages}{739--749}.
\newblock
\showISBNx{978-1-4503-2957-6}
\urldef\tempurl%
\url{https://doi.org/10.1145/2660267.2660271}
\showDOI{\tempurl}


\bibitem[\protect\citeauthoryear{Das, Kramer, Dabbish, and Hong}{Das
  et~al\mbox{.}}{2015}]%
        {das_role_2015}
\bibfield{author}{\bibinfo{person}{Sauvik Das}, \bibinfo{person}{Adam~D.I.
  Kramer}, \bibinfo{person}{Laura~A. Dabbish}, {and} \bibinfo{person}{Jason~I.
  Hong}.} \bibinfo{year}{2015}\natexlab{}.
\newblock \showarticletitle{The {Role} of {Social} {Influence} in {Security}
  {Feature} {Adoption}}. In \bibinfo{booktitle}{\emph{Proceedings of the 18th
  {ACM} {Conference} on {Computer} {Supported} {Cooperative} {Work} \& {Social}
  {Computing}}} \emph{(\bibinfo{series}{{CSCW} '15})}.
  \bibinfo{publisher}{Association for Computing Machinery},
  \bibinfo{address}{Vancouver, BC, Canada}, \bibinfo{pages}{1416--1426}.
\newblock
\showISBNx{978-1-4503-2922-4}
\urldef\tempurl%
\url{https://doi.org/10.1145/2675133.2675225}
\showDOI{\tempurl}


\bibitem[\protect\citeauthoryear{Feal, Calciati, Vallina-Rodriguez, Troncoso,
  and Gorla}{Feal et~al\mbox{.}}{2020}]%
        {feal_angel_2020}
\bibfield{author}{\bibinfo{person}{Álvaro Feal}, \bibinfo{person}{Paolo
  Calciati}, \bibinfo{person}{Narseo Vallina-Rodriguez},
  \bibinfo{person}{Carmela Troncoso}, {and} \bibinfo{person}{Alessandra
  Gorla}.} \bibinfo{year}{2020}\natexlab{}.
\newblock \showarticletitle{Angel or {Devil}? {A} {Privacy} {Study} of {Mobile}
  {Parental} {Control} {Apps}}.
\newblock \bibinfo{journal}{\emph{Proceedings on Privacy Enhancing
  Technologies}} \bibinfo{volume}{2020}, \bibinfo{number}{2}
  (\bibinfo{date}{April} \bibinfo{year}{2020}), \bibinfo{pages}{314--335}.
\newblock
\showISSN{2299-0984}
\urldef\tempurl%
\url{https://doi.org/10.2478/popets-2020-0029}
\showDOI{\tempurl}


\bibitem[\protect\citeauthoryear{Felt, Egelman, and Wagner}{Felt
  et~al\mbox{.}}{2012a}]%
        {felt_ive_2012-1}
\bibfield{author}{\bibinfo{person}{Adrienne~Porter Felt},
  \bibinfo{person}{Serge Egelman}, {and} \bibinfo{person}{David Wagner}.}
  \bibinfo{year}{2012}\natexlab{a}.
\newblock \showarticletitle{I've got 99 problems, but vibration ain't one: a
  survey of smartphone users' concerns}. In
  \bibinfo{booktitle}{\emph{Proceedings of the second {ACM} workshop on
  {Security} and privacy in smartphones and mobile devices}}
  \emph{(\bibinfo{series}{{SPSM} '12})}. \bibinfo{publisher}{Association for
  Computing Machinery}, \bibinfo{address}{New York, NY, USA},
  \bibinfo{pages}{33--44}.
\newblock
\showISBNx{978-1-4503-1666-8}
\urldef\tempurl%
\url{https://doi.org/10.1145/2381934.2381943}
\showDOI{\tempurl}


\bibitem[\protect\citeauthoryear{Felt, Ha, Egelman, Haney, Chin, and
  Wagner}{Felt et~al\mbox{.}}{2012b}]%
        {felt_android_2012}
\bibfield{author}{\bibinfo{person}{Adrienne~Porter Felt},
  \bibinfo{person}{Elizabeth Ha}, \bibinfo{person}{Serge Egelman},
  \bibinfo{person}{Ariel Haney}, \bibinfo{person}{Erika Chin}, {and}
  \bibinfo{person}{David Wagner}.} \bibinfo{year}{2012}\natexlab{b}.
\newblock \showarticletitle{Android permissions: user attention, comprehension,
  and behavior}. In \bibinfo{booktitle}{\emph{Proceedings of the {Eighth}
  {Symposium} on {Usable} {Privacy} and {Security}}}
  \emph{(\bibinfo{series}{{SOUPS} '12})}. \bibinfo{publisher}{Association for
  Computing Machinery}, \bibinfo{address}{New York, NY, USA},
  \bibinfo{pages}{1--14}.
\newblock
\showISBNx{978-1-4503-1532-6}
\urldef\tempurl%
\url{https://doi.org/10.1145/2335356.2335360}
\showDOI{\tempurl}


\bibitem[\protect\citeauthoryear{Ghosh, Badillo-Urquiola, Guha, LaViola~Jr, and
  Wisniewski}{Ghosh et~al\mbox{.}}{2018}]%
        {ghosh_safety_2018}
\bibfield{author}{\bibinfo{person}{Arup~Kumar Ghosh}, \bibinfo{person}{Karla
  Badillo-Urquiola}, \bibinfo{person}{Shion Guha}, \bibinfo{person}{Joseph~J.
  LaViola~Jr}, {and} \bibinfo{person}{Pamela~J. Wisniewski}.}
  \bibinfo{year}{2018}\natexlab{}.
\newblock \showarticletitle{Safety vs. {Surveillance}: {What} {Children} {Have}
  to {Say} about {Mobile} {Apps} for {Parental} {Control}}. In
  \bibinfo{booktitle}{\emph{Proceedings of the 2018 {CHI} {Conference} on
  {Human} {Factors} in {Computing} {Systems}}} \emph{(\bibinfo{series}{{CHI}
  '18})}. \bibinfo{publisher}{Association for Computing Machinery},
  \bibinfo{address}{New York, NY, USA}, \bibinfo{pages}{1--14}.
\newblock
\showISBNx{978-1-4503-5620-6}
\urldef\tempurl%
\url{https://doi.org/10.1145/3173574.3173698}
\showDOI{\tempurl}


\bibitem[\protect\citeauthoryear{Ghosh, Hughes, and Wisniewski}{Ghosh
  et~al\mbox{.}}{2020}]%
        {ghosh_circle_2020}
\bibfield{author}{\bibinfo{person}{Arup~Kumar Ghosh},
  \bibinfo{person}{Charles~E. Hughes}, {and} \bibinfo{person}{Pamela~J.
  Wisniewski}.} \bibinfo{year}{2020}\natexlab{}.
\newblock \showarticletitle{Circle of {Trust}: {A} {New} {Approach} to {Mobile}
  {Online} {Safety} for {Families}}. In \bibinfo{booktitle}{\emph{Proceedings
  of the 2020 {CHI} {Conference} on {Human} {Factors} in {Computing}
  {Systems}}}. \bibinfo{publisher}{ACM}, \bibinfo{address}{Honolulu HI USA},
  \bibinfo{pages}{1--14}.
\newblock
\showISBNx{978-1-4503-6708-0}
\urldef\tempurl%
\url{https://doi.org/10.1145/3313831.3376747}
\showDOI{\tempurl}


\bibitem[\protect\citeauthoryear{Ghosh and Wisniewski}{Ghosh and
  Wisniewski}{2016}]%
        {ghosh_understanding_2016}
\bibfield{author}{\bibinfo{person}{Arup~Kumar Ghosh} {and}
  \bibinfo{person}{Pamela Wisniewski}.} \bibinfo{year}{2016}\natexlab{}.
\newblock \showarticletitle{Understanding {User} {Reviews} of {Adolescent}
  {Mobile} {Safety} {Apps}: {A} {Thematic} {Analysis}}. In
  \bibinfo{booktitle}{\emph{Proceedings of the 19th {International}
  {Conference} on {Supporting} {Group} {Work}}} \emph{(\bibinfo{series}{{GROUP}
  '16})}. \bibinfo{publisher}{Association for Computing Machinery},
  \bibinfo{address}{New York, NY, USA}, \bibinfo{pages}{417--420}.
\newblock
\showISBNx{978-1-4503-4276-6}
\urldef\tempurl%
\url{https://doi.org/10.1145/2957276.2996283}
\showDOI{\tempurl}


\bibitem[\protect\citeauthoryear{Guan, Lee, Cuddihy, and Ramey}{Guan
  et~al\mbox{.}}{2006}]%
        {guan_validity_2006}
\bibfield{author}{\bibinfo{person}{Zhiwei Guan}, \bibinfo{person}{Shirley Lee},
  \bibinfo{person}{Elisabeth Cuddihy}, {and} \bibinfo{person}{Judith Ramey}.}
  \bibinfo{year}{2006}\natexlab{}.
\newblock \showarticletitle{The validity of the stimulated retrospective
  think-aloud method as measured by eye tracking}. In
  \bibinfo{booktitle}{\emph{Proceedings of the {SIGCHI} {Conference} on {Human}
  {Factors} in {Computing} {Systems}}} \emph{(\bibinfo{series}{{CHI} '06})}.
  \bibinfo{publisher}{Association for Computing Machinery},
  \bibinfo{address}{New York, NY, USA}, \bibinfo{pages}{1253--1262}.
\newblock
\showISBNx{978-1-59593-372-0}
\urldef\tempurl%
\url{https://doi.org/10.1145/1124772.1124961}
\showDOI{\tempurl}


\bibitem[\protect\citeauthoryear{Hashish, Bunt, and Young}{Hashish
  et~al\mbox{.}}{2014}]%
        {hashish_involving_2014}
\bibfield{author}{\bibinfo{person}{Yasmeen Hashish}, \bibinfo{person}{Andrea
  Bunt}, {and} \bibinfo{person}{James~E. Young}.}
  \bibinfo{year}{2014}\natexlab{}.
\newblock \showarticletitle{Involving children in content control: a
  collaborative and education-oriented content filtering approach}. In
  \bibinfo{booktitle}{\emph{Proceedings of the {SIGCHI} {Conference} on {Human}
  {Factors} in {Computing} {Systems}}} \emph{(\bibinfo{series}{{CHI} '14})}.
  \bibinfo{publisher}{Association for Computing Machinery},
  \bibinfo{address}{New York, NY, USA}, \bibinfo{pages}{1797--1806}.
\newblock
\showISBNx{978-1-4503-2473-1}
\urldef\tempurl%
\url{https://doi.org/10.1145/2556288.2557128}
\showDOI{\tempurl}


\bibitem[\protect\citeauthoryear{Jambunathan and Counselman}{Jambunathan and
  Counselman}{2002}]%
        {jambunathan_parenting_2002}
\bibfield{author}{\bibinfo{person}{Saigeetha Jambunathan} {and}
  \bibinfo{person}{Kenneth Counselman}.} \bibinfo{year}{2002}\natexlab{}.
\newblock \showarticletitle{Parenting {Attitudes} of {Asian} {Indian} {Mothers}
  {Living} in the {United} {States} and in {India}}.
\newblock \bibinfo{journal}{\emph{Early Child Development and Care}}
  \bibinfo{volume}{172}, \bibinfo{number}{6} (\bibinfo{date}{Dec.}
  \bibinfo{year}{2002}), \bibinfo{pages}{657--662}.
\newblock
\showISSN{0300-4430}
\urldef\tempurl%
\url{https://doi.org/10.1080/03004430215102}
\showDOI{\tempurl}
\newblock
\shownote{Publisher: Routledge \_eprint:
  https://doi.org/10.1080/03004430215102.}


\bibitem[\protect\citeauthoryear{Khovanskaya, Sengers, Mazmanian, and
  Darrah}{Khovanskaya et~al\mbox{.}}{2017}]%
        {khovanskaya_reworking_2017}
\bibfield{author}{\bibinfo{person}{Vera Khovanskaya}, \bibinfo{person}{Phoebe
  Sengers}, \bibinfo{person}{Melissa Mazmanian}, {and} \bibinfo{person}{Charles
  Darrah}.} \bibinfo{year}{2017}\natexlab{}.
\newblock \showarticletitle{Reworking the {Gaps} between {Design} and
  {Ethnography}}. In \bibinfo{booktitle}{\emph{Proceedings of the 2017 {CHI}
  {Conference} on {Human} {Factors} in {Computing} {Systems}}}
  \emph{(\bibinfo{series}{{CHI} '17})}. \bibinfo{publisher}{Association for
  Computing Machinery}, \bibinfo{address}{New York, NY, USA},
  \bibinfo{pages}{5373--5385}.
\newblock
\showISBNx{978-1-4503-4655-9}
\urldef\tempurl%
\url{https://doi.org/10.1145/3025453.3026051}
\showDOI{\tempurl}


\bibitem[\protect\citeauthoryear{Kiesler, Zdaniuk, Lundmark, and Kraut}{Kiesler
  et~al\mbox{.}}{2000}]%
        {kiesler_troubles_2000}
\bibfield{author}{\bibinfo{person}{Sara Kiesler}, \bibinfo{person}{Bozena
  Zdaniuk}, \bibinfo{person}{Vicki Lundmark}, {and} \bibinfo{person}{Robert
  Kraut}.} \bibinfo{year}{2000}\natexlab{}.
\newblock \showarticletitle{Troubles {With} the {Internet}: {The} {Dynamics} of
  {Help} at {Home}}.
\newblock \bibinfo{journal}{\emph{Human–Computer Interaction}}
  \bibinfo{volume}{15}, \bibinfo{number}{4} (\bibinfo{date}{Dec.}
  \bibinfo{year}{2000}), \bibinfo{pages}{323--351}.
\newblock
\showISSN{0737-0024, 1532-7051}
\urldef\tempurl%
\url{https://doi.org/10.1207/S15327051HCI1504_2}
\showDOI{\tempurl}


\bibitem[\protect\citeauthoryear{Kropczynski, Aljallad, Elrod, Lipford, and
  Wisniewski}{Kropczynski et~al\mbox{.}}{2021a}]%
        {kropczynski_towards_2021}
\bibfield{author}{\bibinfo{person}{Jess Kropczynski}, \bibinfo{person}{Zaina
  Aljallad}, \bibinfo{person}{Nathan~Jeffrey Elrod}, \bibinfo{person}{Heather
  Lipford}, {and} \bibinfo{person}{Pamela~J. Wisniewski}.}
  \bibinfo{year}{2021}\natexlab{a}.
\newblock \showarticletitle{Towards {Building} {Community} {Collective}
  {Efficacy} for {Managing} {Digital} {Privacy} and {Security} within {Older}
  {Adult} {Communities}}.
\newblock \bibinfo{journal}{\emph{Proceedings of the ACM on Human-Computer
  Interaction}} \bibinfo{volume}{4}, \bibinfo{number}{CSCW3}
  (\bibinfo{date}{Jan.} \bibinfo{year}{2021}), \bibinfo{pages}{255:1--255:27}.
\newblock
\urldef\tempurl%
\url{https://doi.org/10.1145/3432954}
\showDOI{\tempurl}


\bibitem[\protect\citeauthoryear{Kropczynski, Ghaiumy~Anaraky, Akter, Godfrey,
  Lipford, and Wisniewski}{Kropczynski et~al\mbox{.}}{2021b}]%
        {kropczynski_examining_2021}
\bibfield{author}{\bibinfo{person}{Jess Kropczynski}, \bibinfo{person}{Reza
  Ghaiumy~Anaraky}, \bibinfo{person}{Mamtaj Akter}, \bibinfo{person}{Amy~J.
  Godfrey}, \bibinfo{person}{Heather Lipford}, {and} \bibinfo{person}{Pamela~J.
  Wisniewski}.} \bibinfo{year}{2021}\natexlab{b}.
\newblock \showarticletitle{Examining Collaborative Support for Privacy and
  Security in the Broader Context of Tech Caregiving}.
\newblock \bibinfo{journal}{\emph{Proc. ACM Hum.-Comput. Interact.}}
  \bibinfo{volume}{5}, \bibinfo{number}{CSCW2}, Article
  \bibinfo{articleno}{396} (\bibinfo{date}{oct} \bibinfo{year}{2021}),
  \bibinfo{numpages}{23}~pages.
\newblock
\urldef\tempurl%
\url{https://doi.org/10.1145/3479540}
\showDOI{\tempurl}


\bibitem[\protect\citeauthoryear{Livingstone, Haddon, Goerzig, and
  Ólafsson}{Livingstone et~al\mbox{.}}{2011}]%
        {livingstone_risks_2011}
\bibfield{author}{\bibinfo{person}{Sonia Livingstone}, \bibinfo{person}{Leslie
  Haddon}, \bibinfo{person}{Anke Goerzig}, {and} \bibinfo{person}{Kjartan
  Ólafsson}.} \bibinfo{year}{2011}\natexlab{}.
\newblock \showarticletitle{Risks and {Safety} on the {Internet}: {The}
  {Perspective} of {European} {Children}. {Full} {FINDINGS}}.
\newblock  (\bibinfo{date}{Jan.} \bibinfo{year}{2011}).
\newblock


\bibitem[\protect\citeauthoryear{Madigan, Ly, Rash, Van~Ouytsel, and
  Temple}{Madigan et~al\mbox{.}}{2018}]%
        {madigan_prevalence_2018}
\bibfield{author}{\bibinfo{person}{Sheri Madigan}, \bibinfo{person}{Anh Ly},
  \bibinfo{person}{Christina~L. Rash}, \bibinfo{person}{Joris Van~Ouytsel},
  {and} \bibinfo{person}{Jeff~R. Temple}.} \bibinfo{year}{2018}\natexlab{}.
\newblock \showarticletitle{Prevalence of {Multiple} {Forms} of {Sexting}
  {Behavior} {Among} {Youth}: {A} {Systematic} {Review} and {Meta}-analysis}.
\newblock \bibinfo{journal}{\emph{JAMA pediatrics}} \bibinfo{volume}{172},
  \bibinfo{number}{4} (\bibinfo{year}{2018}), \bibinfo{pages}{327--335}.
\newblock
\showISSN{2168-6211}
\urldef\tempurl%
\url{https://doi.org/10.1001/jamapediatrics.2017.5314}
\showDOI{\tempurl}


\bibitem[\protect\citeauthoryear{Mendel and Toch}{Mendel and Toch}{2017}]%
        {mendel_susceptibility_2017}
\bibfield{author}{\bibinfo{person}{Tamir Mendel} {and} \bibinfo{person}{Eran
  Toch}.} \bibinfo{year}{2017}\natexlab{}.
\newblock \bibinfo{title}{Susceptibility to {Social} {Influence} of {Privacy}
  {Behaviors} {\textbar} {Proceedings} of the 2017 {ACM} {Conference} on
  {Computer} {Supported} {Cooperative} {Work} and {Social} {Computing}}.
\newblock
\newblock
\urldef\tempurl%
\url{https://dl.acm.org/doi/10.1145/2998181.2998323}
\showURL{%
\tempurl}


\bibitem[\protect\citeauthoryear{Mendel and Toch}{Mendel and Toch}{2019}]%
        {mendel_my_2019}
\bibfield{author}{\bibinfo{person}{Tamir Mendel} {and} \bibinfo{person}{Eran
  Toch}.} \bibinfo{year}{2019}\natexlab{}.
\newblock \showarticletitle{My {Mom} was {Getting} this {Popup}:
  {Understanding} {Motivations} and {Processes} in {Helping} {Older}
  {Relatives} with {Mobile} {Security} and {Privacy}}.
\newblock \bibinfo{journal}{\emph{Proceedings of the ACM on Interactive,
  Mobile, Wearable and Ubiquitous Technologies}} \bibinfo{volume}{3},
  \bibinfo{number}{4} (\bibinfo{date}{Dec.} \bibinfo{year}{2019}),
  \bibinfo{pages}{147:1--147:20}.
\newblock
\urldef\tempurl%
\url{https://doi.org/10.1145/3369821}
\showDOI{\tempurl}


\bibitem[\protect\citeauthoryear{Mitchell, Jones, Finkelhor, and
  Wolak}{Mitchell et~al\mbox{.}}{2014}]%
        {mitchell_trends_2014}
\bibfield{author}{\bibinfo{person}{Kimberly~J. Mitchell},
  \bibinfo{person}{Lisa~M. Jones}, \bibinfo{person}{David Finkelhor}, {and}
  \bibinfo{person}{Janis Wolak}.} \bibinfo{year}{2014}\natexlab{}.
\newblock \showarticletitle{Trends in {Unwanted} {Online} {Experiences} and
  {Sexting} : {Final} {Report}}. \bibinfo{address}{Durham, NH: Crimes against
  Children Research Center}.
\newblock


\bibitem[\protect\citeauthoryear{Moser, Chen, and Schoenebeck}{Moser
  et~al\mbox{.}}{2017}]%
        {moser_parents_2017}
\bibfield{author}{\bibinfo{person}{Carol Moser}, \bibinfo{person}{Tianying
  Chen}, {and} \bibinfo{person}{Sarita~Y. Schoenebeck}.}
  \bibinfo{year}{2017}\natexlab{}.
\newblock \showarticletitle{Parents? and {Children}?s {Preferences} about
  {Parents} {Sharing} about {Children} on {Social} {Media}}. In
  \bibinfo{booktitle}{\emph{Proceedings of the 2017 {CHI} {Conference} on
  {Human} {Factors} in {Computing} {Systems}}}. \bibinfo{publisher}{ACM},
  \bibinfo{address}{Denver Colorado USA}, \bibinfo{pages}{5221--5225}.
\newblock
\showISBNx{978-1-4503-4655-9}
\urldef\tempurl%
\url{https://doi.org/10.1145/3025453.3025587}
\showDOI{\tempurl}


\bibitem[\protect\citeauthoryear{NW, 800Washington, and Inquiries}{NW
  et~al\mbox{.}}{2013}]%
        {nw_teens_2013}
\bibfield{author}{\bibinfo{person}{1615 L.~St NW}, \bibinfo{person}{Suite
  800Washington}, {and} \bibinfo{person}{DC~20036USA202-419-4300 {\textbar}
  Main202-857-8562 {\textbar} Fax202-419-4372 {\textbar}~Media Inquiries}.}
  \bibinfo{year}{2013}\natexlab{}.
\newblock \bibinfo{title}{Teens and {Mobile} {Apps} {Privacy}}.
\newblock
\newblock
\urldef\tempurl%
\url{https://www.pewresearch.org/internet/2013/08/22/teens-and-mobile-apps-privacy/}
\showURL{%
\tempurl}


\bibitem[\protect\citeauthoryear{NW, 800Washington, and Inquiries}{NW
  et~al\mbox{.}}{2015}]%
        {nw_teens_2015-1}
\bibfield{author}{\bibinfo{person}{1615 L.~St NW}, \bibinfo{person}{Suite
  800Washington}, {and} \bibinfo{person}{DC~20036USA202-419-4300 {\textbar}
  Main202-857-8562 {\textbar} Fax202-419-4372 {\textbar}~Media Inquiries}.}
  \bibinfo{year}{2015}\natexlab{}.
\newblock \bibinfo{title}{Teens, {Technology} and {Romantic} {Relationships}}.
\newblock
\newblock
\urldef\tempurl%
\url{https://www.pewresearch.org/internet/2015/10/01/teens-technology-and-romantic-relationships/}
\showURL{%
\tempurl}


\bibitem[\protect\citeauthoryear{NW, 800Washington, and Inquiries}{NW
  et~al\mbox{.}}{2018}]%
        {nw_majority_2018}
\bibfield{author}{\bibinfo{person}{1615 L.~St NW}, \bibinfo{person}{Suite
  800Washington}, {and} \bibinfo{person}{DC~20036USA202-419-4300 {\textbar}
  Main202-857-8562 {\textbar} Fax202-419-4372 {\textbar}~Media Inquiries}.}
  \bibinfo{year}{2018}\natexlab{}.
\newblock \bibinfo{title}{A {Majority} of {Teens} {Have} {Experienced} {Some}
  {Form} of {Cyberbullying}}.
\newblock
\newblock
\urldef\tempurl%
\url{https://www.pewresearch.org/internet/2018/09/27/a-majority-of-teens-have-experienced-some-form-of-cyberbullying/}
\showURL{%
\tempurl}


\bibitem[\protect\citeauthoryear{Pain}{Pain}{2006}]%
        {pain_paranoid_2006}
\bibfield{author}{\bibinfo{person}{Rachel Pain}.}
  \bibinfo{year}{2006}\natexlab{}.
\newblock \showarticletitle{Paranoid parenting? {Rematerializing} risk and fear
  for children}.
\newblock \bibinfo{journal}{\emph{Social \& Cultural Geography}}
  \bibinfo{volume}{7}, \bibinfo{number}{2} (\bibinfo{date}{April}
  \bibinfo{year}{2006}), \bibinfo{pages}{221--243}.
\newblock
\showISSN{1464-9365}
\urldef\tempurl%
\url{https://doi.org/10.1080/14649360600600585}
\showDOI{\tempurl}


\bibitem[\protect\citeauthoryear{Rader and Wash}{Rader and Wash}{2015}]%
        {rader_identifying_2015}
\bibfield{author}{\bibinfo{person}{Emilee Rader} {and} \bibinfo{person}{Rick
  Wash}.} \bibinfo{year}{2015}\natexlab{}.
\newblock \showarticletitle{Identifying patterns in informal sources of
  security information}.
\newblock \bibinfo{journal}{\emph{Journal of Cybersecurity}}
  \bibinfo{volume}{1}, \bibinfo{number}{1} (\bibinfo{date}{Sept.}
  \bibinfo{year}{2015}), \bibinfo{pages}{121--144}.
\newblock
\showISSN{2057-2085}
\urldef\tempurl%
\url{https://doi.org/10.1093/cybsec/tyv008}
\showDOI{\tempurl}
\newblock
\shownote{Publisher: Oxford Academic.}


\bibitem[\protect\citeauthoryear{Rader, Wash, and Brooks}{Rader
  et~al\mbox{.}}{2012}]%
        {rader_stories_2012}
\bibfield{author}{\bibinfo{person}{Emilee Rader}, \bibinfo{person}{Rick Wash},
  {and} \bibinfo{person}{Brandon Brooks}.} \bibinfo{year}{2012}\natexlab{}.
\newblock \showarticletitle{Stories as informal lessons about security}. In
  \bibinfo{booktitle}{\emph{Proceedings of the {Eighth} {Symposium} on {Usable}
  {Privacy} and {Security}}} \emph{(\bibinfo{series}{{SOUPS} '12})}.
  \bibinfo{publisher}{Association for Computing Machinery},
  \bibinfo{address}{Washington, D.C.}, \bibinfo{pages}{1--17}.
\newblock
\showISBNx{978-1-4503-1532-6}
\urldef\tempurl%
\url{https://doi.org/10.1145/2335356.2335364}
\showDOI{\tempurl}


\bibitem[\protect\citeauthoryear{Reardon, Feal, Wijesekera, On,
  Vallina-Rodriguez, and Egelman}{Reardon et~al\mbox{.}}{2019}]%
        {reardon_50_2019}
\bibfield{author}{\bibinfo{person}{Joel Reardon}, \bibinfo{person}{Álvaro
  Feal}, \bibinfo{person}{Primal Wijesekera}, \bibinfo{person}{Amit Elazari~Bar
  On}, \bibinfo{person}{Narseo Vallina-Rodriguez}, {and} \bibinfo{person}{Serge
  Egelman}.} \bibinfo{year}{2019}\natexlab{}.
\newblock \showarticletitle{50 {Ways} to {Leak} {Your} {Data}: {An}
  {Exploration} of {Apps}' {Circumvention} of the {Android} {Permissions}
  {System}}. \bibinfo{pages}{603--620}.
\newblock
\showISBNx{978-1-939133-06-9}
\urldef\tempurl%
\url{https://www.usenix.org/conference/usenixsecurity19/presentation/reardon}
\showURL{%
\tempurl}


\bibitem[\protect\citeauthoryear{Steinberg, Lamborn, Dornbusch, and
  Darling}{Steinberg et~al\mbox{.}}{1992}]%
        {steinberg_impact_1992}
\bibfield{author}{\bibinfo{person}{Laurence Steinberg},
  \bibinfo{person}{Susie~D. Lamborn}, \bibinfo{person}{Sanford~M. Dornbusch},
  {and} \bibinfo{person}{Nancy Darling}.} \bibinfo{year}{1992}\natexlab{}.
\newblock \showarticletitle{Impact of {Parenting} {Practices} on {Adolescent}
  {Achievement}: {Authoritative} {Parenting}, {School} {Involvement}, and
  {Encouragement} to {Succeed}}.
\newblock \bibinfo{journal}{\emph{Child Development}} \bibinfo{volume}{63},
  \bibinfo{number}{5} (\bibinfo{year}{1992}), \bibinfo{pages}{1266--1281}.
\newblock
\showISSN{0009-3920}
\urldef\tempurl%
\url{https://doi.org/10.2307/1131532}
\showDOI{\tempurl}
\newblock
\shownote{Publisher: [Wiley, Society for Research in Child Development].}


\bibitem[\protect\citeauthoryear{Van~Parys}{Van~Parys}{2019}]%
        {works_you_2019}
\bibfield{author}{\bibinfo{person}{Bill Van~Parys}.}
  \bibinfo{year}{2019}\natexlab{}.
\newblock \bibinfo{title}{You don’t need to be tech savvy to be a tech
  caregiver}.
\newblock
\newblock
\urldef\tempurl%
\url{https://www.fastcompany.com/90438110/you-dont-need-to-be-tech-savvy-to-be-a-tech-caregiver}
\showURL{%
\tempurl}


\bibitem[\protect\citeauthoryear{Vogels and Anderson}{Vogels and
  Anderson}{2019}]%
        {vogels_americans_2019}
\bibfield{author}{\bibinfo{person}{Emily~A. Vogels} {and}
  \bibinfo{person}{Monica Anderson}.} \bibinfo{year}{2019}\natexlab{}.
\newblock \showarticletitle{Americans and {Digital} {Knowledge}}.
\newblock \bibinfo{journal}{\emph{Pew Research}} (\bibinfo{date}{Oct.}
  \bibinfo{year}{2019}).
\newblock
\urldef\tempurl%
\url{https://www.pewresearch.org/internet/2019/10/09/americans-and-digital-knowledge/}
\showURL{%
\tempurl}


\bibitem[\protect\citeauthoryear{Wang, Zhao, Van~Kleek, and Shadbolt}{Wang
  et~al\mbox{.}}{2021}]%
        {wang2021a}
\bibfield{author}{\bibinfo{person}{G Wang}, \bibinfo{person}{J Zhao},
  \bibinfo{person}{M Van~Kleek}, {and} \bibinfo{person}{N Shadbolt}.}
  \bibinfo{year}{2021}\natexlab{}.
\newblock \showarticletitle{Protection or punishment? relating the design space
  of parental control apps and perceptions about them to support parenting for
  online safety}.
\newblock \bibinfo{journal}{\emph{Proceedings of the Conference on Computer
  Supported Cooperative Work Conference}}.
\newblock


\bibitem[\protect\citeauthoryear{Wisniewski, Ghosh, Xu, Rosson, and
  Carroll}{Wisniewski et~al\mbox{.}}{2017}]%
        {wisniewski_parental_2017}
\bibfield{author}{\bibinfo{person}{Pamela Wisniewski},
  \bibinfo{person}{Arup~Kumar Ghosh}, \bibinfo{person}{Heng Xu},
  \bibinfo{person}{Mary~Beth Rosson}, {and} \bibinfo{person}{John~M. Carroll}.}
  \bibinfo{year}{2017}\natexlab{}.
\newblock \showarticletitle{Parental {Control} vs. {Teen} {Self}-{Regulation}:
  {Is} there a middle ground for mobile online safety?}. In
  \bibinfo{booktitle}{\emph{Proceedings of the 2017 {ACM} {Conference} on
  {Computer} {Supported} {Cooperative} {Work} and {Social} {Computing}}}.
  \bibinfo{publisher}{ACM}, \bibinfo{address}{Portland Oregon USA},
  \bibinfo{pages}{51--69}.
\newblock
\showISBNx{978-1-4503-4335-0}
\urldef\tempurl%
\url{https://doi.org/10.1145/2998181.2998352}
\showDOI{\tempurl}


\end{thebibliography}
Received July 2021; revised November 2021; accepted November 2021.
\newpage


\begin{appendices}
\section{}


\setcounter{table}{0} 
\renewcommand{\thetable}{A.\arabic{table}}
\begin{table*}[h]
 \centering
 \footnotesize
 \caption{Structure of Interview with Sample Questions}
   \label{tab:interview-questions}
\begin{tabular}{ |p{3cm}|p{9cm}|  }
 \hline
\textbf{Structure} & \textbf{Sample Questions} \\ \hline
\textbf{RQ1:} Current approaches for managing mobile online safety and privacy &  
\begin{itemize}[leftmargin=0.2cm]
   \item  \textit{How do you decide which apps are safe or unsafe to install on your mobile phone?}
   \item \textit{How do you decide whether a permission request is safe to accept or deny?}
   \item  \textit{As a family, do you discuss with one another how to make good decisions about what apps to install or what permissions to accept?}
   \item \textit{Do you monitor your family's phones to see what apps they are using?}
   \item \textit{Who in your family would you consider most tech savvy? Who do you go to if you have a concern about the privacy and safety of apps installed on your phone?}
 \end{itemize} \\ \hline

\textbf{RQ2:} Evaluation of app features designed for co-managing mobile privacy and online safety & 

\textbf{Task 1:} Select which of the installed apps you are willing to hide from your family members. 
\begin{itemize}[leftmargin=0.4cm]
   \item  \textit{Are there any apps in your phone that you would not want to show to some of your family members?}
   \item \textit{Do you think a teen/parent should be able to hide their apps from their parents/children?}
   \item  \textit{What are some of the benefits and drawbacks of allowing users to hide some of their apps from family members?}
 \end{itemize} 
 
\textbf{Task 2:} Review the apps and permissions of your teen’s/parent’s device and identify any that you think might be a concern.
\begin{itemize}[leftmargin=0.4cm]
   \item \textit{Do you see any apps that your teen/parent has installed that might be a concern?} 
   \item \textit{Do you see any granted permission on your teen/parent’s phone that could be a privacy concern?}
 \end{itemize} 
  
\textbf{Task 3:} Send a message to warn your parent/teen about this app.
\begin{itemize}[leftmargin=0.4cm]
   \item \textit{If you saw a concerning app or permission on your teen’s phone, would you use this message feature to contact them to warn about it?} 
   \item \textit{If someone in your family privately message you and inform you about any of your apps or permissions, would this influence you to uninstall or change your permission in any way?}
 \end{itemize} 
 
 \textbf{Task 4:} Post a comment about the concerning app in the Community Page.
\begin{itemize}[leftmargin=0.4cm]
   \item \textit{When sharing a concern about an app, what information do you think is important to share?}
   \item \textit{If one of your family members posted a negative review of an app, would this influence you to uninstall the app in any way?}
 \end{itemize} 
 \\ \hline

\textbf{RQ3:} Considerations in designing an app for co-managing mobile privacy and online safety for families &  \begin{itemize}[leftmargin=0.2cm]
  \item \textit{What were some of the features of the app that you liked and disliked? Why?}
   \item  \textit{Would you use an app like this in your family? Why or Why not?} 
     \item \textit{How could this app be improved to work for you and your family members?}
     \item \textit{Would you use this app more to check your own apps and permissions or those of your family members’?}
   \item  \textit{Who do you think in your family would use this app? For instance, would it make sense to use the app with your extended family members (e.g., grandparents, aunts, uncles, cousins)?}
 \end{itemize} \\ \hline


\end{tabular}
\end{table*}
\newpage
\section{}

\setcounter{table}{0} 
\renewcommand{\thetable}{B.\arabic{table}}
\begin{table*}[h]
  \centering
  \footnotesize
\caption{Codebook for RQ1}
  \label{tab:codebook-RQ1}
 \begin{tabular}{| >{\raggedright}m{2.5cm} | >{\raggedright}m{2.8cm}  | m{7.2cm} |} 
\hline
 \textbf{Theme} & \textbf{Parent/Teen Codes} & \textbf{Illustrative Quotations} \\ 
 \hline
 
 \multicolumn{3}{c}{} \\[-10pt]  
 \multirow{3}{2.5cm}{Most parents and teens install apps with little consideration about privacy and online safety.} & No Special\newline (P: 63\%, N=12) \newline (T: 53\%, N=10) & \textit{"Usually what I do is the requirement basis that if I need the functionality of an app, I go ahead and install."} – P19 \\ \cline{2-3}
 
 & Do Some Research \newline(P: 32\%, N=6) \newline (T: 37\%, N=7) & \textit{“I usually try to look at their [App's] reviews to be honest, and do a little bit of background research on the app. So yeah. Just look it up on Google.”} -T2   \\ \cline{2-3}
   
 & Ask Teens/Parents \newline(P: 5\%, N=1) \newline (T: 26\%, N=5) & \textit{“Yes, I usually when I want to install an app, I have to ask her first. And there have been apps that I've been told I cannot install.”} – T2   \\ \cline{1-3}
 
 \multicolumn{3}{c}{} \\[-10pt] 
 \multirow{4}{2.5cm}{ Most parents and teens accepted permissions for their apps to function properly.} & Accept If Required \newline(P: 79\%, N=15) \newline (T: 89\%, N=17) & \textit{“There are some apps that if you don't accept all permissions, it won't work properly, so either I may just choose not to use that one or just decide that I'm okay with the things its asking to me”}–P15   \\ \cline{2-3}
 
 & Accept All \newline(P: 16\%, N=3) \newline (T: 11\%, N=2) & \textit{“Well, I don’t use that many apps. They are just games and so, I accept.”}  -T4   \\ \cline{2-3}

 & Reject All \newline(P: 5\%, N=1)   & \textit{“Usually if it ever gives me an alert like that, I just hit deny.”}-P4   \\ \hline
 
 \multicolumn{3}{c}{} \\[-10pt] 
 \multirow{4}{2.5cm}{Although teens provided general tech support to their parents, parents often manually check their teens’ app usage.} &  No Checks \newline(P: 26\%, N=5) \newline (T: 100\%, N=19) & \textit{“No, I don’t because they are parents. They have freedom.”} - T11 \\ \cline{2-3}
 
 & Manually Looked \newline(P: 47\%, N=9)  & \textit{“Well, we would take the phone [Teen's] and I will check to see if there is an app that we don’t know about.”} – P16  \\ \cline{2-3}
   
 & Parental Control Apps \newline(P: 26\%, N=5)  & \textit{“We've always enabled parental control on our [Teen's] apps. So if she does download an app, we automatically get notified.”}-P3   \\ \cline{2-3}
 
 & Provides Tech Support \newline(P: 26\%, N=5) \newline (T: 74\%, N=14) & \textit{“He is 16 and he knows about computer. He [Teen] knows more than me. He shows me how to use computers.”} -P5   \\ [1ex] 
 \hline  
\end{tabular}
\end{table*}
\newpage
\section{}

\setcounter{table}{0} 
\renewcommand{\thetable}{C.\arabic{table}}
\begin{table*}[h]
  \centering
  \footnotesize
\caption{Codebook for RQ2}
  \label{tab:codebook-RQ2}
 \begin{tabular}{| >{\raggedright}m{2.5cm} | >{\raggedright}m{3cm}  | m{7cm} |} 
\hline
 \textbf{Themes (Features)} & \textbf{Parent/Teen Codes} & \textbf{Illustrative Quotations} \\ 
 \hline

 \multicolumn{3}{c}{} \\[-10pt]  
 \multirow{3}{2.5cm}{Parents and Teens valued the features for Collaborative privacy and online safety Management.} &  Ensures Family Safety \newline (P: 63\%, N=12) \newline (T: 74\%, N=14) & \textit{“it's at a glance, you can just help each other to stay safe as a family.”-P15} \\ \cline{2-3}
 
  & Allows Family Oversight \newline (P: 68\%, N=13) \newline (T: 42\%, N=8) & \textit{“It’s nice to have somebody going through all the apps permissions and pointing out some of the permissions that not so okay”-P19}   \\ \cline{2-3}
   
  & Promotes Awareness \newline (P: 37\%, N=7)  & \textit{“I like it actually, this and the creating this awareness that they [Teens] should not download and give permission, like as a blank check to all their apps.”-P12}  \\ \hline
 
 \multicolumn{3}{c}{} \\[-10pt]  
 \multirow{4}{2.5cm}{The communication features were not overall appreciated among families. Parents preferred talking instead of messaging.} & Would Talk Instead \newline(P: 58\%, N=11) \newline (T: 26\%, N=5) & \textit{“ it's much easier to go and talk to her [Teen] directly rather than sending a message.
 ” – P19} \\ \cline{2-3}
 
 & Use When Apart \newline(P: 37\%, N=7) \newline (T: 21\%, N=4) & \textit{“If she's [Teen] not available to talk, talk immediately. So, I can just send out a message so that I don't, we didn't forget that issue.”-P19}   \\ \cline{2-3}
 
 & Use Other Tools Instead\newline (P: 21\%, N=4)\newline (T: 32\%, N=6) & \textit{"Probably not, because we already have a like text message group. So I'll just maybe writing that it there."-T8}  \\ \cline{2-3}

 & Positive: Promotes Discussion\newline (P: 26\%, N=5)  \newline (T: 21\%, N=4)  & \textit{“The involvement of each family member in the decision making of the utilization of the app, I think that's the best feature that it has. Again, it brings a family to make the decisions that are going to be better for the entire family.”-P14}  \\ \hline

\multicolumn{3}{c}{} \\[-10pt]  
 \multirow{5}{2.5cm}{Both Parents and Teens did not approve of the features that decreased transparency.} & Promotes Risky Behavior \newline(P: 84\%, N=16)  \\ (T: 32\%, N=6)  & \textit{“The more open you are in terms of the apps that you have, the safer you are going to be, because it means accountability. Right?
 ”-P14} \\ \cline{2-3}
 
  & Affects Transparency \newline (P: 42\%, N=8) \newline (T: 53\%, N=10) & \textit{“if they have something, I feel like they should feel confident enough to talk to their parents with about everything.”-P18}   \\ \cline{2-3}

  & Privacy Not Needed \newline (P: 26\%, N=5) \newline (T: 42\%, N=8)  & \textit{“there's nothing really that I want to hide on my phone.” -T15}  \\ \cline{2-3}
 
 &  Promotes Distrust \newline (T: 26\%, N=5)  & \textit{“Well, I guess since you know, that you're able to hide apps, from them [Parents] so they might want to check it out anyway. See, you can't hide it anyway.” -T15}   \\ \cline{2-3}

 & Positive: Allows Personal Privacy \newline (P: 37\%, N=7) \newline (T: 47\%, N=9) & \textit{“you could also have your privacy for apps that you know, you don't want people to know about.”-T16}     \\ [1ex] \hline
 

   
\end{tabular}
\end{table*}

\newpage
\section{}

\setcounter{table}{0} 
\renewcommand{\thetable}{D.\arabic{table}}
\begin{center}
 \footnotesize
\begin{longtable}{| >{\raggedright}m{2.5cm} | >{\raggedright}m{3cm}  | m{6.5cm} |} 
\caption{Codebook for RQ3} \label{tab:codebook-RQ3} \\

\hline \multicolumn{1}{|c|}{\textbf{Themes}} & \multicolumn{1}{c|}{\textbf{Parent/Teen Codes}} & \multicolumn{1}{c|}{\textbf{Illustrative Quotations}} \\ \hline 
\endfirsthead

\multicolumn{3}{c}%
{{\tablename\ \thetable{} -- continued from previous page}} \\
\hline \multicolumn{1}{|c|}{\textbf{Themes}} & \multicolumn{1}{c|}{\textbf{Parent/Teen Codes}} & \multicolumn{1}{c|}{\textbf{Illustrative Quotations}} \\ \hline 
\endhead

\hline \multicolumn{3}{|r|}{{Continued on next page}} \\ \hline
\endfoot
\hline
\endlastfoot
 
 \multicolumn{3}{c}{} \\[-10pt] 
 \multicolumn{1}{|c|}{} & \multicolumn{2}{c|}{App Level}  \\ \cline{2-3}
 \multirow{5}{2.4cm}[15pt]{Parents were more concerned about teens’ apps usage while teens were more focused on the granted permissions of parents’ phones.} & Did Not Find \newline Concerning App \newline (P: 26\%, N=5) \newline (T: 79\%, N=15) & \textit{"I mean, he's [Teen] very vigilant on what he downloads and doesn't. So I don't really see anything that's bad."-P6}  \\ \cline{2-3}

 & Found Concerning \newline App - Unknown \newline (P: 63\%, N=12) \newline (T: 11\%, N=2)  & \textit{“Why do you [Teen] have a McMaster app? What is it McMaster Carr? I don't even know what this is.  I need to check with you later”-P6}    \\ \cline{2-3}
   
 & Found Concerning \newline App - Social \newline(P: 21\%, N=4)  \newline(T: 11\%, N=2) & \textit{“She has Instagram because one of her school programs wanted her to have this account. but this might have some problems because if she keeps accepting unknown people's invitations."-P7}    \\ \cline{2-3}
  
& Found Concerning \newline App - Financial \newline(P: 16\%, N=3) & \textit{“I see that he has the Venmo app. So I, I have no idea why he would need the Venmo. He's not paying anybody."-P8}    \\ \cline{2-3} 
    
 
\multicolumn{1}{|c|}{} & \multicolumn{2}{c|}{Permission Level}  \\ \cline{2-3}
 & Did Not Find \newline Concerning Permission \newline(P: 68\%, N=13) \newline (T: 42\%, N=8) & \textit{“ I don't see any concerning as of now with just after exploring"-P10}     \\ \cline{2-3}
 
   & Found Concerning Permission - Location \newline(P: 26\%, N=5) \newline (T: 32\%, N=6) & \textit{“Prime video has location granted. That is, that is little questionable.”-T19} \\ \cline{2-3}

  & Found Concerning Permission - Others \newline(P: 5\%, N=1) \newline (T: 26\%, N=5) & \textit{“His WhatsApp, he granted the camera and his accounts to it."-T14}   \\ \hline
 \multicolumn{3}{c}{} \\[-10pt] 
 \multirow{8}{2.4cm}[15pt]{Parents would monitor their teens’ and other family member’s privacy and online safety more where teens would mostly monitor their own apps.} &  Would Monitor Only Own Apps \newline (P: 37\%, N=7) \newline (T: 74\%, N=14) & \textit{“Um, I don't need um, as a kid, I really don't think it matters. What apps my parents are using, that's not really my concern. I would probably just use it to check my app."-T11}   \\ \cline{2-3}

 & Would Monitor One Another \newline (P: 63\%, N=12)  \newline (T: 26\%, N=5)  & \textit{“As parents, youre just concerned of whatever it is coming to the life of your children. So you're going to be looking at from that perspective, and so using it [CO-oPS] even more.”–P14}  \\ \cline{2-3}
 
  & Would Monitor All in Family \newline (P: 58\%, N=11) \newline(T: 11\%, N=2) & \textit{“I would use the app [CO-oPS] just, I could see the apps and permissions granted, and then tell my family so that they are also aware. So I think that's what will be beneficial the most.
 "-P16}   \\ \hline

 \multicolumn{3}{c}{} \\[-10pt]  
 \multirow{3}{2.5cm}[15pt]{Parents would listen to their teen’s oversight, but teens need to verify first.} & Would Change App Settings \newline(P: 89\%, N=17) \newline (T: 26\%, N=5) & \textit{"I think I would definitely go in and change the permission If somebody sent it to me, then I would say, yeah, let me know about it, and then I'll change the permission."}-P3  \\ \cline{2-3}
 
 & Verify Before Changing \newline (P: 11\%, N=2) \newline (T: 68\%, N=13) & \textit{“Yes, if they make some good points, Because there could be some moments of miscommunication where one party thinks the app is damaging, and the other party knows it isn't in some cases.”}-T1    \\ \cline{2-3}
   
 & Would Not Change \newline (T: 5\%, N=1) &  \textit{“If someone points me at something, I don’t think I would need to change my apps, because I know what apps I use and their permissions.”}-T6   \\ \hline
 \multicolumn{3}{c}{} \\[-10pt]  
 \multirow{6}{2.5cm}[15pt]{Where Teens liked to be treated as peers in co-monitoring with their parents’ app safety, but parents wanted more power.} & Liked Equal Co-Monitoring Power \newline  (P: 21\%, N=4) \newline (T: 64\%, N=12) & \textit{“I really like that, because the teen could actually feel like they are on the same level as parent.”-T17}  \\ \cline{2-3}
   
 & Wanted Control on Teen’s/Parent's Settings   \newline (P: 32\%, N=6) \newline (T: 16\%, N=3) & \textit{"I could see what was there and what was happening, but I can't change it. So my gut reaction was like, Oh, I don't like that."}-P9  \\ \cline{2-3}

& Wanted Notification on Teen’s/Parent's App Changes \newline (P: 32\%, N=6) \newline (T: 11\%, N=2) & \textit{"I didn't like it [Co-managing feature]. So, then if my daughter installs a new app, that there can be pop up notifications on my side. So that that would be very useful."}-P7  \\ \cline{2-3}

& Disliked being Treated as Peers (P: 26\%, N=5)  & \textit{“But so far, it looks like we're peers in and I would want to be able to control a little more."}-P2   \\ \cline{2-3}

& Disliked- Teens Having Privacy \newline (P: N=13, 68\%)  & \textit{“I would prefer that for a kid, they not be able to hide the apps on their phones. And maybe it would be like another layer where the parents have that option”}-P2   \\ \cline{2-3}
 
& Wanted Control on Teen’s Privacy \newline (P: N=6, 32\%) & \textit{“the parent has to have the option to allow them [Teens] to hide [the apps they have installed].”}-P9   \\ \cline{2-3}


& No Comment \newline (T: 11\%, N=2) & \\

\end{longtable}
\end{center}

\end{appendices}
\end{document}